\documentclass[aps,prb,twocolumn,english,showpacs,showkeys]{revtex4-1}

\usepackage{dcolumn}
\usepackage{bm}

\usepackage{graphicx} 
\usepackage{epsfig}
\usepackage{amsfonts}
\usepackage[figures right]{rotating}
\usepackage{amssymb}
\usepackage{amsmath}
\usepackage{amsmath}
\usepackage{graphicx}
\usepackage{hyperref}
\usepackage{color}
\usepackage{soul}
\usepackage{tabularx}
\usepackage{array}
\usepackage{epstopdf}  
\usepackage{verbatim}

\newcommand\numberthis{\addtocounter{equation}{1}\tag{\theequation}}

\begin{document}
\title{Thermal conductivity of local moment models with strong spin-orbit coupling}

\author{Georgios L. Stamokostas}
\email{georgios@physics.utexas.edu}
\affiliation{Department of Physics, The University of Texas at Austin, Austin, TX, 78712, USA} 
\author{Panteleimon E. Lapas}
\thanks{G.S. and P.L. contributed equally to this work}
\affiliation{Department of Physics, The University of Texas at Austin, Austin, TX, 78712, USA}
\author{Gregory A. Fiete}
\affiliation{Department of Physics, The University of Texas at Austin, Austin, TX, 78712, USA}

\date{\today}

\begin{abstract}
We study the magnetic and lattice contributions to the thermal conductivity of electrically insulating strongly spin-orbit coupled magnetically ordered phases on a two-dimensional honeycomb lattice using the Kitaev-Heisenberg model. Depending on model parameters, such as the relative strength of the spin-orbit induced anisotropic coupling, a number of magnetically ordered phases are possible. In this work, we study two distinct regimes of thermal transport depending on whether the characteristic energy of the phonons or the magnons dominates, and focus on two different relaxation mechanisms, boundary scattering and magnon-phonon scattering. For spatially anisotropic magnetic phases, the thermal conductivity tensor can be highly anisotropic when the magnetic energy scale dominates, since the magnetic degrees of freedom dominate the thermal transport for temperatures well below the magnetic transition temperature. In the opposite limit in which the phonon energy scale dominates, the thermal conductivity will be nearly isotropic, reflecting the isotropic (at low temperatures) phonon dispersion assumed for the honeycomb lattice. We further discuss the extent to which thermal transport properties are influenced by strong spin-orbit induced anisotropic coupling in the local moment regime of insulating magnetic phases. The developed methodology can be applied to any 2D magnon-phonon system, and more importantly to systems where an analytical Bogoliubov transformation cannot be found and magnon bands are not necessarily isotropic.
\end{abstract}


\maketitle


\section{\label{sec:intro}Introduction}

In recent years, the intense research activity around topological insulators \cite{Moore:nat10,Hasan:rmp10,Qi:rmp11,Ando:jpsj13} has drawn increased attention to the influence of spin-orbit coupling in the solid state, and demonstrated that qualitatively new phases of band insulators can appear.\cite{Bernevig:prl06,Bernevig:sci06,Kane:prl05,Kane_2:prl05}  In the limit of strong electron-electron interactions, spin-orbit coupling can also have a profound influence on the phase diagram of Hamiltonians potentially relevant to correlated topological materials.\cite{Maciejko:np15}  In the context of correlated materials with strong spin-orbit coupling, transition metal oxides containing iridium atoms, known as iridates, have been a focus of research.\cite{Kargarian:prl13,Kargarian:prb11,Pesin:np10,Shitade:prl09,Krempa:arcm14,Go:prl12,Witczak:prb12,Wan:prb11,Rau:arcm16,Schaffer:cm15}   In particular, unusual magnetic orders have been suggested in a number of iridates.\cite{Rau:prl14,Kargarian:prb12,Reuther:prb12,Chaloupka:prl13,Jackeli:prl09,Chaloupka:prl10,Katukuri:njp14,Kim_Ir:prb14,Sizyuk:prb14,Rau:prl14,Yamaji:prl14,Katukuri:prx14,Hozoi:prb14,Mazin:prl12,Jiang:prb11,Kimchi:prb15,Lee:prb16,Rousochatzakis:prx15,Cook:prb15,Knolle:prl14}  Due to the large cross-section for neutron absorption in iridium, neutron scattering experiments (which can be used to determine magnetic order as well as the magnon spectra) are especially challenging,\cite{Choi:prl12} and a variety of experimental techniques have been applied to study them.\cite{Singh:prl12,Comin:prl12,Singh:prl12,Gretarsson:prl13,Trousselet:prl13,Alpichshev:prl15,Takayama:prl15}   In particular, resonant X-ray scattering is a powerful probe of iridates.\cite{Moretti:prl14,Ament:rmp11,Kim_Ir:prb14,Kotani:rmp01,Williams:prb16,Biffin:prb14}  The magnetic model we study in this work is in part motivated by theoretical work on iridates.

Another area in which spin-orbit coupling has played a leading role is spintronics,\cite{Dietl:rmp14,Tserkovnyak:rmp05,Zutic:rmp04} where the coupling of spin and orbital motion allows for an electrical detection of spin properties. Spintronic devices offer the possibility of low-power components of computing elements, and may also exhibit longer coherence times than conventional devices, which may prove useful for quantum architectures.\cite{Hanson:rmp07}   In spin caloritronic devices the additional element of a thermal gradient is included and the relationship between thermal gradients, spin currents, and spin-orbit induced voltages is investigated.\cite{Adachi:rpp13,Bauer:natnm12,Uchida2014}  

In this work, we are interested in the thermal transport properties of a 2D strong spin-orbit driven magnetic insulator.  In these systems, the thermal transport is dominated (at low temperatures) by magnetic and lattice excitations that carry heat. We study these systems using local moment models that are coupled to phonons (lattice distortions) through exchange constants that depend on the relative distance between nearby moments.  The main role of the spin-orbit coupling is to induce unusual, and sometimes highly  spatially anisotropic magnetic orders.  The thermal transport is computed within the Boltzmann approach (in the relaxation time approximation) which takes as inputs the magnon spectrum of the various magnetic orders, and the phonon spectrum of the underlying lattice.  For concreteness, we focus on a well-known two-dimensional model, the so-called Heisbenberg-Kitaev (HK) model,\cite{Chaloupka:prl10,Chaloupka:prl13,Singh:prl12,Jiang:prb11} on the honeycomb lattice.  The HK model has a rich, established magnetic phase diagram that provides a useful starting point for investigating the magnetic fluctuations within the $1/S$ expansion, where $S$ is the magnitude of the local moment.\cite{Auerbach}  Previous studies of thermal transport in insulating magnetic materials indicated that the magnetic and thermal contributions to the thermal conductivity can be comparable.\cite{Dixon:prb80,Li:prl05,Mikhail:pbcm11,Mikhail:pbcm15,Chernyshev:prb15,Chernyshev:prl16}  Our main result in this work is to show that the spatially anisotropic magnetic states that can arise from strong spin-orbit coupling can dramatically affect the thermal transport, or have a rather small effect depending on the relative size of magnon and phonon thermal conductivities.   In some cases, the thermal transport may help identify the symmetries of the magnetically ordered state if other measurements are difficult or problematic. 

This paper is organized as follows.  In Sec.\ref{sec:magnon-phonon} we introduce the local moment model we study, and describe how the phonons are incorporated into the exchange constants of the model.  In Sec.\ref{sec:spin_waves} the magnon spectrum for various ordered phases of the local moment model is computed, which will be used as an input for the thermal conductivity.  In Sec.\ref{sec:mag_pho_relxtimes} the magnon and phonon scattering rates are computed, and in Sec.\ref{sec:thermal_conductivity} we present the results for the thermal conductivity in various regimes and for various phases of our model.  Finally, we present the main conclusions in Sec.\ref{sec:conclusions}.  Several lengthy technical details are relegated to the appendices.

\section{\label{sec:magnon-phonon} Model Hamiltonian}
\begin{figure}[h]
    \includegraphics[scale=0.7]{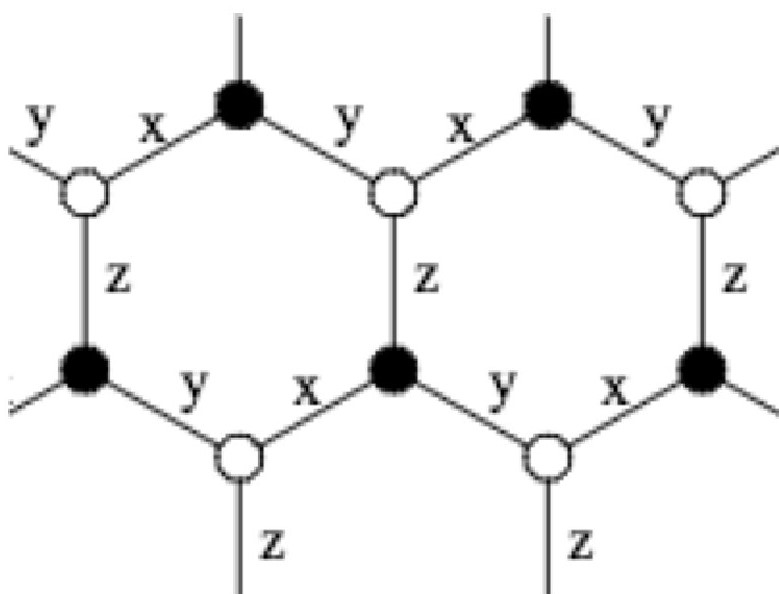}
    \caption{Honeycomb lattice with bond labels, $\gamma  = \{ x,y,z\} $, used for the Kitaev terms in Eq.\eqref{eq:HKham}.}
    \label{fig:Kitaevbonds}
 \end{figure}

We consider a total Hamiltonian for local moments coupled to the lattice as $ \cal {\hat H} = {\cal \hat H}^{\rm spin} + {\cal \hat H}^{\rm pho}$, where the coupling between spin and lattice (phonon) degrees of freedom will be made explicit below.  We study a local moment model with an established phase diagram, the Heisenberg-Kitaev (HK) model defined on a two-dimensional honeycomb lattice with  nearest neighbor (NN) interactions:\cite{Chaloupka:prl10,Chaloupka:prl13,Singh:prl12,Jiang:prb11}
\begin{equation}
\label{eq:HKham}
\hat {\cal H}^{\rm spin}=\sum_{\langle i, j\rangle}\hat H_{ij}^{(\gamma )} = \sum_{\langle i, j\rangle} \Big({J_{ij}}{{\bf{S}}_i} \cdot {{\bf{S}}_j}+ 2{K_{ij}}S_i^\gamma S_j^\gamma\Big) ,
\end{equation}
where $\gamma  = \{ x,y,z\} $   labels the three distinct types of NN bonds, as shown in Fig.\ref{fig:Kitaevbonds}, $i$ and $j$ label sites of the lattice, and $S_i^\gamma$ is the $\gamma^{th}$ component of the local moment on site $i$.  The first term in Eq.\eqref{eq:HKham} describes a rotationally invariant (in spin space) Heisenberg interaction between nearest-neighbor spins and the second term is the so-called ``Kitaev" term \cite{Kitaev2006} that describes bond-direction-dependent anisotropic spin interactions.  One may view it as originating from an underlying spin-orbit coupling.\cite{Kargarian:prb12}  The exchange constants, $J_{ij}$ and $K_{ij}$ describe the relative strengths of the Heisenberg and Kitaev terms respectively.

\begin{figure}[h]
    \centering
    \includegraphics[scale=0.3]{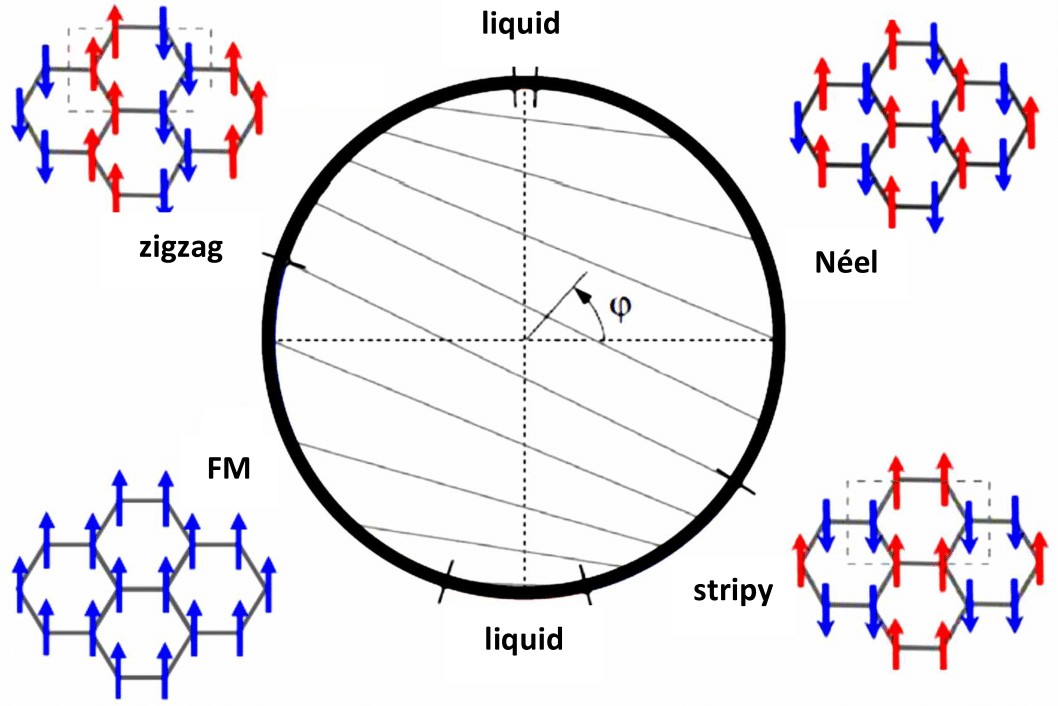}
    \caption{(Color online) Phase diagram of the Kitaev-Heisen-berg model with the parametrization of Eq.\eqref{eq:HK_A}. A variety of magnetic and non-magnetic ``liquid" phases are present as a function of the angle $\varphi$.\cite{Chaloupka:prl13}  A schematic of the various ordered states is shown. The magnetic unit cell for the zigzag and the stripy phase is shown as a dashed rectangle.}
    \label{fig:phasediagKH}
\end{figure}

The HK Hamiltonian of Eq.\eqref{eq:HKham},  using $A=\sqrt{K^2+J^2}$ (where $K$ and $J$ are the magnitudes of the nearest nei-ghbor Kitaev and Heisenberg exchange couplings), can be expressed in terms of a parameter  $\varphi$ such that $K=A\sin\varphi$, $J=A\cos\varphi$ and $\varphi \in [0,2\pi]$, as\cite{Chaloupka:prl13}
 \begin{equation}
\label{eq:HK_A}
\hat {\cal H}^{\rm spin} = \sum_{\langle i, j\rangle} A({\cos\varphi}{{\bf{S}}_i} \cdot {{\bf{S}}_j}+ 2{\sin\varphi}S_i^\gamma S_j^\gamma ),
\end{equation}
and its phase diagram is shown in Fig.\ref{fig:phasediagKH}. For fixed $A$ (which sets an overall energy scale), there are a wide range of magnetic (and non-magnetic spin-liquid) phases.  In this work, we focus on the ferromagnetic, N\'eel, stripy, and zig-zag phases. The presence of the Kitaev coupli-ngs additionally renders the low energy magnetic excitations of the various magnetically ordered phases spatially anisotropic, as a result of which the thermal conductivity, especially if it is magnon dominated, is generally expected to be different ``along" the stripe (or zig-zag) compared to the direction ``perpendicular" to it. 

In this work, we are interested in the heat carried by both magnetic and lattice degrees of freedom.  We consider only temperatures lower than the Debye temperature, and retain the energy of the lattice displacements to quadratic order to obtain a phonon spectrum in the standard way.\cite{AM_book}  The generic resulting phonon Hamiltonian (in the absence of coupling to magnons) in second quantized form is given by
 \begin{equation}
 \label{eq:Hphonon}
 {\cal \hat{H}}^{\rm pho}= \sum_{{\bf{q}},s} \hbar \omega_{{\bf{q}}s}  c_{{\bm{q}}s} ^\dagger c_{{\bm{q}}s},
 \end{equation} 
 where $s$ labels the type of phonon polarization, $c^\dagger_{\bm{q}s}$ ($c_{\bm{q}s}$)  the creation (annihilation) operator of a phonon of wavevector $\bm{q}$  and polarization $s$, and $\omega_{\bm{q}s}$ is its eigenfrequency.  At temperatures much lower than the Debye temperature, we can use the Debye model for acoustic phonons (that are of interest in this work), which assumes that $\omega_{{\bm{q}}s} =v |{\bm q}|$ , i.e. the phonon dispersion is spatially isotropic.  We further assume the phonons are two-dimensional, and therefore they only disperse within the plane of the honeycomb lattice.

The coupling between the phonons and the magnons enters through the distance dependence of the exchange constants, $J_{ij}=J(\bm{r}_i-\bm{r}_j)$ , $K_{ij}=K(\bm{r}_i-\bm{r}_j)$, where ${\bm r}_i$ and ${\bm r}_j$ denote the dynamic position of the ions at the $i^{th}$ and $j^{th}$ lattice sites.  Assuming a small displacement of the ions from their equilibrium positions (long phonon wavelength approximation consistent also with the linear isotropic phonon dispersion given above), the exchange constants can be approximated as\cite{Dixon:prb80}
 \begin{equation}
 \label{eq:jexpand}
 {J_{ij}} = J({\bm{R}}_i^{} + {\bm{u}}_i^{} - {\bm{R}}_j^{} - {{\bm{u}}_j})= J({\bm{R}}_{ij}^{}) + {\bm{u}}_{ij}^{} \cdot {\bm{J}}_{ij}'+ ...,
 \end{equation}
 \begin{equation}
  \label{eq:kexpand}
  {K_{ij}} = K({\bm{R}}_i^{} + {\bm{u}}_i^{} - {\bm{R}}_j^{} - {{\bm{u}}_j})= K({\bm{R}}_{ij}^{}) + {\bm{u}}_{ij}^{} \cdot {\bm{K}}_{ij}'+ ..., 
\end{equation}
 where
\begin{equation}
   {\bm{J}}_{ij}'={{\left. {{\nabla _{{{\bm{r}}_{ij}}}}J({{\bm{r}}_{ij}})} \right|}_{{{\bm{r}}_{ij}} = {{\bm{R}}_{ij}}}},\;
  {\bm{K}}_{ij}'={{\left. {{\nabla _{{{\bm{r}}_{ij}}}}K({{\bm{r}}_{ij}})} \right|}_{{{\bm{r}}_{ij}} = {{\bm{R}}_{ij}}}}, \nonumber
  \end{equation}
 are gradients with respect to $\bm{r}_{ij} $ evaluated at the equilibrium magnetic ion distances $\bm{R}_{ij}$.
Here, ${{\bm{u}}_{ij}} \equiv {\bm{u}}_i^{} - {{\bm{u}}_j}$, ${{\bm{r}}_{ij}} \equiv {\bm{r}}_i^{} - {{\bm{r}}_j} = {\bm{R}}_i^{} + {{\bm{u}}_i} - {\bm{R}}_j^{} - {{\bm{u}}_j}$, and 
${\bm{R}}_{ij}^{} \equiv {\bm{R}}_i^{} - {\bm{R}}_j^{}$.  

The ionic displacement from its equilibrium position is expressed in terms of phonon creation and annihilation operators as\cite{Mahan,Solyom} 
\begin{equation}
\label{eq:latticedisplacoper}
{{\bm{u}}_{i\tau}} = \sum\limits_{{\bm{q}},s } {\sqrt {\frac{\hbar }{{2NM{\omega _{{\bm{q}}s }}}}} \left( {c_{ - {\bm{q}}s }^\dag  + c_{{\bm{q}}s }^{}} \right)} {e^{i{\bm{q}} \cdot {{\bm{R}}_i}}} {{{\bm{\hat e}}}_{{\bm{q}}s \tau }},
\end{equation}
where $N$ is the total number of chemical unit cells, $M$ the mass of the magnetic atoms (assumed of the same type on each sublattice), and ${{\bm{\hat e}}_{{\bm{q}}s \tau }}$ the direction of the displacement of the magnetic ion at the $i^{th}$ lattice position of the $\tau^{th}$ sublattice (the honeycomb lattice has two sublattice sites), relative to a phonon of polarization $s$  and direction of propagation given by  $\bm{q}$.

Within the long wavelength approximation valid for the acoustic phonons, ionic displacements from their equilibrium positions are taken sublattice independent, and denoted as 
\begin{equation}
\label{eq:ftlatticedisplacoper}
{{\bm{u}}_{i}} =\frac{1}{\sqrt{N}} \sum_{\bm{q}} {e^{i{\bm{q}} \cdot {{\bm{R}}_i}}}\vec{u}_{\bm{q}},
\end{equation}
where $\bm{R}_i$ denotes the lattice equilibrium position of a magnetic ion, and we have defined 
\begin{equation}
\label{eq:uq}
{\vec{u}_{\bm{q}}} = \sum\limits_{s } {\sqrt {\frac{\hbar }{{2M{\omega _{{\bm{q}}s }}}}}       \left( {c_{ - {\bm{q}}s }^\dag  + c_{{\bm{q}}s }^{}} \right)}      {{{\bm{\hat e}}}_{{\bm{q}}s  }}.
\end{equation}
Substituting Eqs.\eqref{eq:jexpand} and \eqref{eq:kexpand} into Eq.\eqref{eq:HKham},  one finds an expansion of the magnetic part of the total Hamiltonian in powers of phonon operators,
\begin{equation}
\label{eq:hexpand}
{\cal \hat H}_{\rm spin}^{\rm pho} = {\cal \hat H}_{\rm spin}^{0\rm pho} + {\cal \hat H}_{\rm spin}^{1 \rm pho} + {\cal \hat H}_{\rm spin}^{2 \rm pho} + ...
\end{equation}
where the first term is the spin-phonon Hamiltonian with magnetic ions at their lattice equilibrium positions, the second term is the coupling of one power of phonon operators with the spin system, the third term the coupling of two powers of phonon operators with the spin system and so on. 

In the low temperature regime and under the assumption of weak magnon-phonon coupling,  one-phonon processes are more important than multiple phonon processes, and therefore we truncate the infinite expansion of Eq.\eqref{eq:hexpand} up to the ${\cal \hat H}_{\rm spin}^{1 \rm pho} $ term. More specifically, we retain the following two terms of Eq.\eqref{eq:hexpand},
\begin{equation}
{\cal \hat H}_{\rm spin}^{0\rm pho} =\sum\limits_{\left\langle {ij} \right\rangle } {J({\bm{R}}_{ij}^{})\;{{\bm{S}}_i} \cdot {{\bm{S}}_j}}  + \sum\limits_{\left\langle {ij} \right\rangle } {2K({\bm{R}}_{ij}^{})\;S_i^\gamma S_j^\gamma },
\end{equation}
\begin{equation}
\label{eq:H_pert}
{\cal \hat H}_{\rm spin}^{1\rm pho}=\sum\limits_{\left\langle {ij} \right\rangle } {\left( {{\bm{u}}_{ij}^{} \cdot {\bm{J}}_{ij}'} \right)  {{\bm{S}}_i} \cdot {{\bm{S}}_j}}  + 2\sum\limits_{\left\langle {ij} \right\rangle } {\left( {{\bm{u}}_{ij}^{} \cdot {\bm{K}}_{ij}'} \right)\;S_i^\gamma S_j^\gamma }.
\end{equation}
In our Boltzmann approach to the thermal transport, Eq.\eqref{eq:H_pert} will be treated perturbatively as a term that scatters magnons and phonons, leading to a finite lifetime (and scattering rate) of each.

\section{\label{sec:spin_waves} Magnons and Scattering Amplitudes}
The phase diagram of Eq.\eqref{eq:HKham} and its extension, Eq.\eqref{eq:HK_A}, has been obtained previously in the literature.\cite{Chaloupka:prl10,Chaloupka:prl13,Singh:prl12,Jiang:prb11}  Here, we are interested in the magnetic excitations above the ground state, which are needed to compute the thermal transport due to the magnetic degrees of freedom.  To the best of our knowledge, only for some phases of the nn HK model have the magnon spectra been previously obtained \cite{Plakida:jetp16}.

We compute the magnon spectrum by representing the three Hermitian spin operators $\bm{S}_i=(S_i^x,S_i^y,S_i^z)$ with Bose operators using the Holstein-Primakoff (HP) representation\cite{Auerbach} (see below) which employs a Taylor expansion in powers of $1/S$ in the spin operators around the classical ground state, as a result of which the ${\cal \hat H}_{\rm spin}^{0\rm pho}$  and ${\cal \hat H}_{\rm spin}^{1\rm pho}$  terms are decomposed as 
\begin{equation}
\label{eq:eqspinexpansion}
{\cal \hat H}_{\rm spin}^{0\rm pho} = {\cal \hat H}_{0\rm mag}^{0\rm pho} + {\cal \hat H}_{1\rm mag}^{0\rm pho} + {\cal \hat H}_{2\rm mag}^{0\rm pho} + ...
\end{equation}
\begin{equation}
\label{eq:phspinexpansion}
{\cal \hat H}_{\rm spin}^{1\rm pho} = {\cal \hat H}_{0\rm mag}^{1\rm pho} + {\cal \hat H}_{1\rm mag}^{1\rm pho} + {\cal \hat H}_{2\rm mag}^{1\rm pho} + ...
\end{equation}
In Eq.\eqref{eq:eqspinexpansion} the first term represents a classical spin background, and the rest of the terms are one magnon, two magnon (and so on) terms. In Eq.\eqref{eq:phspinexpansion},  the first term represents the propagation of one phonon in a classical spin background, the second term the coexistence of one phonon and one magnon (that for non-collinear phases leads to magnon-phonon hybridization), the third term the coexistence of one phonon and two magnons and so on. 

At temperatures much lower than the magnetic transition temperature (which we assume throughout our analysis), the linear spin wave approximation for the magnon energies can be used. The terms trilinear, quadrilinear and higher order in the magnon operators lead to a renormalization of the magnon bands via magnon-magnon interactions in  Eq.\eqref{eq:eqspinexpansion}, and are assumed to be negligible in the low-temperature limit. Furthermore, due to the smallness of the magnon and phonon populations in the temperature regime of interest, we similarly discard terms of higher order in the magnon operators in Eq.\eqref{eq:phspinexpansion}. We further note that the ${\cal \hat H}_{1\rm mag}^{0\rm pho}$ and ${\cal \hat H}_{1\rm mag}^{1\rm pho}$ terms are zero for collinear magnetic orders (all the magnetic orders in Fig.\ref{fig:phasediagKH} are collinear), which can be seen straightforwardly by using the HP representation in the the linear spin wave approximation.  Therefore, the remaining dominant interaction term is ${\cal \hat H}_{2\rm mag}^{1\rm pho}$.

Since the magnetic phase diagram of our spin Hamiltonian includes only collinear states, we define a positive $z$-direction (choice is arbitrary) for the ordered moments, and in the linear spin wave approximation, local moments that are in the positive direction are expanded as 
\begin{eqnarray}
\label{eq:positiveHPoperators1}
S_i^{||} &=& S - a_i^\dag a_i, \\
\label{eq:positiveHPoperators2}
S_i^ +  &\approx& \sqrt {2S} a_i, \\
\label{eq:positiveHPoperators3}
S_i^ -  &\approx& \sqrt {2S} a_i^\dag, 
\end{eqnarray}
while  local moments that lie in the opposite direction are expanded as 
\begin{eqnarray}
\label{eq:negativeHPoperators1}
S_i^{||} &=& -S + b_i^\dag b_i, \\
\label{eq:negativeHPoperators2}
S_i^ +  &\approx& \sqrt {2S} b_i^\dag, \\
\label{eq:negativeHPoperators3}
S_i^ -  &\approx& \sqrt {2S} b_i, 
\end{eqnarray}
where  $a_i^\dagger a_i$ creates a spin deviation of the local moment that lies along the positive z-direction and is located at the $i^{th}$ lattice position, at the $a$-sublattice, and correspondingly  for $b_j^\dagger b_j$, which refers to a local moment aligned along the negative z-direction.  We can switch to a $\bm{k}$-space (momentum space) representation by using the following Fourier transform conventions
\begin{eqnarray}
\label{Fourieroperators1}
&{a_i } = \sqrt {\frac{4}{N}} \sum\limits_{\bm{k}}^{} {{e^{i\vec k \cdot {{\vec \alpha }_0}}}} {a_{{\bm{k}}}},                       a_i ^\dag  = \sqrt {\frac{4}{N}} \sum\limits_{\bm{k}}^{} {{e^{ - i\vec k \cdot {{\vec \alpha }_0}}}}  a_{{\bm{k}}}^\dag , \\
\label{Fourieroperators2} 
&{b_j } = \sqrt {\frac{4}{N}} \sum\limits_{\bm{k}}^{} {{e^{i\vec k \cdot {{\vec \beta }_0}}}} {b_{{\bm{k}}}},                       b_j ^\dag  = \sqrt {\frac{4}{N}} \sum\limits_{\bm{k}}^{} {{e^{ - i\vec k \cdot {{\vec \beta }_0}}}}  b_{{\bm{k}}}^\dag, 
\end{eqnarray}
where $\vec{\alpha}_0$, $\vec{\beta}_0$ are the equilibrium positions of the magnetic ions on the $a^{th}$ and $b^{th}$ sublattice, and we take into account the fact that we have four magnetic sublattices for the stripy and the zig-zag phase, each of $N/4$ magnetic ions, and two magnetic sublattices for the N\'eel and the ferromagnetic phase, each of $N/2$ magnetic ions (in which case the prefactor in Eqs. \eqref{Fourieroperators1} and \eqref{Fourieroperators2} is $\sqrt{2/N}$), given that the N\'eel and the ferromagnetic phase have a magnetic unit cell that is the same as the chemical unit cell of the honeycomb lattice whereas the magnetic unit cell of the stripy and the zig-zag phase is twice the size of the chemical unit cell of the honeycomb lattice. 

Our total Hamiltonian ${\cal \hat H}={\cal \hat H}_0+{\cal \hat H}_{\rm int}$ decomposes into the non-interacting part ${\cal \hat H}_0={\cal \hat H}_{0\rm mag}^{0\rm pho}+{\cal \hat H}_{2\rm mag}^{0\rm pho}+{\cal \hat H}^{\rm pho}$, and the lowest order interacting term ${\cal \hat H_{\rm int}}={\cal \hat H}_{2\rm mag}^{1\rm pho}$,
where ${\cal \hat H}^{\rm pho}$ is given by Eq.\eqref{eq:Hphonon}, ${\cal \hat H}_{0\rm mag}^{1\rm pho}=0$ (from the conventional phonon theory), and ${\cal \hat H}_{0\rm mag}^{0\rm pho} = {\cal H}_{classical}$. The non-diagonal two magnon part of ${\cal \hat H}_0$  in the compact Nambu representation (that takes into account four magnetic sublattices) is given by
\begin{equation}
\label{eq:spinwave}
{{\cal \hat H}_{2\rm mag}^{0\rm pho}}  =  \frac{S}{2}\sum\limits_{\bm{k}} {{\Psi ^\dag }({\bm{k}}){M} ({\bm{k}})} \Psi ({\bm{k}}),
\end{equation}
The sum in Eq.\eqref{eq:spinwave} extends over all wavevectors $\bm{k}$ in the first magnetic Brillouin zone, and by definition it is ${\Psi ^\dag }({\bm{k}}) = \left[ {\begin{array}{*{20}{c}}
  {a_{\bm{k}}^\dag }&{b_{\bm{k}}^\dag }&{c_{\bm{k}}^\dag }&{d_{\bm{k}}^\dag }&{a_{ - {\bm{k}}}^{}}&{b_{ - {\bm{k}}}^{}}&{c_{ - {\bm{k}}}^{}}&{d_{ - {\bm{k}}}^{}} 
\end{array}} \right]$, with $a_{\bm{k}}^\dag$    ($a_{\bm{k}}$) creating (annihilating) a plane-wave magnon mode on sublattice $a$ and so on,
and ${M} ({\bm{k}})$ is an 8$\times$8 (or 4$\times$4 in the case of the N\'eel and ferromagnetic phases) matrix containing information about the spin wave modes of each magnetic phase (see Appendix~\ref{sec:magnon_spectrum}). 

In the same magnon operator representation, the interacting Hamiltonian for the one phonon-two magnon processes is written as
\begin{equation}
\label{eq:interactionmatrix}
{{\cal \hat H}_{2\rm mag}^{1\rm pho}}=\frac{S}{{2\sqrt N }} {\sum\limits_{\bm{k},\bm{q}} {{\Psi ^\dag }({\bm{k}}){\Lambda } ({\bm{k}},{\bm{q}})} } \Psi (\bm{k}-\bm{q}),
\end{equation}
for phonons with wavevector $\bm{q}$ and magnons with wave-vectors $\bm{k}$, and $\bm{k}-\bm{q}$ respectively, where   momentum conservation has been taken into account, and $\Lambda(\bm{k,q})$ is an 8$\times$8 (or 4$\times$4 for the N\'eel and ferromagnetic phases) matrix that contains information about the magnon-phonon interaction (it encompasses the gradient terms appearing in Eqs.\eqref{eq:jexpand} and \eqref{eq:kexpand}).
To switch from the non-diagonal Hamiltonian $\frac{S}{2}\sum\limits_{\bm{k}} {{\Psi ^\dag }({\bm{k}}){M} ({\bm{k}})} \Psi ({\bm{k}})$ of Eq.\eqref{eq:spinwave} to a diagonal one that uses non-interacting magnon modes, we symbolically introduce a Bogoliubov-Valatin transformation,\cite{Ming-wen-Xiao2011}
\begin{equation}
\label{eq:bogoltransform}
\Psi ({\bm{k}}) = {U} ({\bm{k}})\Phi ({\bm{k}}),
\end{equation}
where 
\begin{equation}
\label{eq:diagbase}
  {\Phi ^\dag }({\bm{k}}) = \left[ {\begin{array}{*{20}{c}}
  {\alpha _{\bm{k}}^\dag }&{\beta _{\bm{k}}^\dag }&{\gamma _{\bm{k}}^\dag }&{\delta _{\bm{k}}^\dag }&{\alpha _{ - {\bm{k}}}^{}}&{\beta _{ - {\bm{k}}}^{}}&{\gamma _{ - {\bm{k}}}^{}}&{\delta _{ - {\bm{k}}}^{}} 
\end{array}} \right].
\end{equation}
The 8$ \times $8 (or 4$\times$4 in the case of the N\'eel and ferromagnetic phases) coefficient matrix
${U} ({\bm{k}})$  of Eq.\eqref{eq:bogoltransform} satisfies the following properties   
for all momenta in the first magnetic Brillouin zone,
\[\begin{gathered}
  {{{U} }^\dag }({\bm{k}}){M} ({\bm{k}}){{{U} }^{}}({\bm{k}}) =\\
   Diag\left\{ {{\omega _1}({\bm{k}}),...,{\omega _4}({\bm{k}}), - {\omega _5}({\bm{k}}),..., - {\omega _8}({\bm{k}})} \right\},   \\ 
   \end{gathered}\]
  where 
${\omega _5} =  - {\omega _1},\;{\omega _6} =  - {\omega _2},\;{\omega _7} =  - {\omega _3},\;{\omega _8} =  - {\omega _4}$, and
\[\begin{gathered}  
  {{{U} }^\dag }({\bm{k}})\;{I_ - }\;{U} ({\bm{k}}) = I,\\
  {I_ - } = \left[ {\begin{array}{*{20}{c}}
  {{I_{4 \times 4}}}&{{0_{4 \times 4}}} \\ 
  {{0_{4 \times 4}}}&{ - {I_{4 \times 4}}} 
\end{array}} \right],\; I = \left[ {\begin{array}{*{20}{c}}
  {{I_{4 \times 4}}}&{{0_{4 \times 4}}} \\ 
  {{0_{4 \times 4}}}&{{I_{4 \times 4}}} 
\end{array}} \right].   \\ 
\end{gathered} \]
That is,  $U({\bm{k}})$ acts as a unitary transformation that  diagonalizes the $M$-matrix, and it also preserves the bosonic nature of the magnon operators.

Under the symbolic Bogoliubov-Valatin transformation of Eq.\eqref{eq:bogoltransform} the ${\cal \hat H}_{2\rm mag}^{0\rm pho}$ term becomes
\begin{align*}
 &{\cal \hat H}_{2\rm mag}^{0\rm pho} = \frac{S}{2}\sum\limits_{\bm{k}} {{\Psi ^\dag }({\bm{k}}){M} ({\bm{k}})} \Psi ({\bm{k}}) =\\ 
   &\frac{S}{2}\sum\limits_{\bm{k}} {{\Phi ^\dag }({\bm{k}}){{{U} }^\dag }({\bm{k}}){M} ({\bm{k}}){U} ({\bm{k}})\Phi ({\bm{k}})}  =   \\ 
   &S\sum\limits_{\bm{k}} \bigg( {\omega _1}({\bm{k}})\alpha _{\bm{k}}^\dag \alpha _{\bm{k}}^{} + {\omega _2}({\bm{k}})\beta _{\bm{k}}^\dag \beta _{\bm{k}}^{} + {\omega _3}({\bm{k}})\gamma _{\bm{k}}^\dag \gamma _{\bm{k}}^{} \\ 
   &+ {\omega _4}({\bm{k}})\delta _{\bm{k}}^\dag \delta _{\bm{k}}^{} + {\omega _1}({\bm{k}})\alpha _{ - {\bm{k}}}^\dag \alpha _{ - {\bm{k}}}^{} + {\omega _2}({\bm{k}})\beta _{ - {\bm{k}}}^\dag \beta _{ - {\bm{k}}}^{} \\ 
   &+ {\omega _3}({\bm{k}})\gamma _{ - {\bm{k}}}^\dag \gamma _{ - {\bm{k}}}^{} + {\omega _4}({\bm{k}})\delta _{ - {\bm{k}}}^\dag \delta _{ - {\bm{k}}}^{}  \bigg), \numberthis 
\end{align*}
where ${\omega _i}({\bm{k}}),\;i = \{ 1,...,8\} $  (or $i = \{1,...,4\}$ for the N\'eel and ferromagnetic phases) are the solutions of the secular equation
$\left| {{D} ({\bm{k}}) - \omega ({\bm{k}}){I}}\right| = 0$, in which 
\begin{equation}
\label{eq:dynamicalmatrix}
{D} ({\bm{k}})=I_-{M}(\bm{k})
\end{equation}
 is the so called dynamical matrix of Ref.[\onlinecite{Ming-wen-Xiao2011}].

The multiplication with the $I_-$ matrix is necessary in order to preserve the Bose commutation relations for the new magnon operators.  We mention just for comparison that in the case of Fermi systems,  where anticommutation relations are used, this is not necessary because the latter are satisfied automatically, and the dynamical matrix for fermions is equal to the $M$-matrix, rendering the diagonalization process easier since $M$ is always a Hermitian matrix (as is the original Hamiltonian), while the dynamical matrix is not guaranteed to be Hermitian in all cases since it differs from the original Hamiltonian.\citep{Ming-wen-Xiao2011}
 The unitary transformation ${U} ({\bm{k}})$  is constructed  by taking the eigenvectors of the dynamical matrix $[\upsilon(\omega_i(\bm{k}))]_{1 \times 8}$ and using them as column vectors as below,\citep{Ming-wen-Xiao2011}
 \begin{equation}
 {U} ({\bm{k}})=[\upsilon(\omega_1(\bm{k})),...,\upsilon(\omega_4(\bm{k})),\upsilon(\omega_5(\bm{k})),...,\upsilon(\omega_8(\bm{k}))]_{8 \times 8}.
 \end{equation}

We next express the interaction Hamiltonian ${\cal \hat H}_{\rm int}$ in terms of the new magnon quasiparticle operators (by  applying the Bogoliubov-Valatin transformation of Eq.\eqref{eq:bogoltransform}) as
\begin{widetext}
\begin{eqnarray}
\label{eq:H1phon2mag}
{\cal \hat H}_{2\rm mag}^{1\rm pho} &=& \frac{S}{2 \sqrt{N}}\sum\limits_{\bm{k},\bm{q},s}  \sqrt {\frac{\hbar }{{2M{\omega _{{\bm{q}}s }}}}} \Bigg(c_{ - {\bm{q}}s}^\dag {\Psi ^\dag }({\bm{k}}){\Lambda}' ({\bm{k}},{\bm{q}})\Psi (\bm{k}-\bm{q}) + c_{{\bm{q}}s}^{}{\Psi ^\dag }({\bm{k}}){\Lambda}' ({\bm{k}},{\bm{q}})\Psi (\bm{k}-\bm{q})\Bigg)\\ \nonumber & =&\frac{S}{2 \sqrt{N}} \sum\limits_{\bm{k},\bm{q},s} \sqrt {\frac{\hbar }{{2M{\omega _{{\bm{q}}s }}}}}\Bigg(c_{ - {\bm{q}}s}^\dag {\Phi ^\dag }({\bm{k}}){U^\dag} ({\bm{k}}){\Lambda}' ({\bm{k}},{\bm{q}}){U} (\bm{k}-\bm{q})\Phi (\bm{k}-\bm{q})
 + c_{{\bm{q}}s}^{}{\Phi ^\dag }({\bm{k}}){U^\dag} (\bm{k}){\Lambda}' ({\bm{k}},{\bm{q}}){U} (\bm{k}-\bm{q})\Phi (\bm{k}-\bm{q}) \Bigg),
 \end{eqnarray} 
 \end{widetext}
where the matrix $\Lambda '(\bm{k},\bm{q})$ and the matrix $\Lambda (\bm{k},\bm{q})$ of Eq.\eqref{eq:interactionmatrix} are related as
\begin{equation}
\label{eq:Lambdap}
\Lambda(\bm{k},\bm{q}) = \sum\limits_{s } {\sqrt {\frac{\hbar }{{2M{\omega _{{\bm{q}}s }}}}}       \left( {c_{ - {\bm{q}}s }^\dag  + c_{{\bm{q}}s }^{}} \right)} \Lambda '(\bm{k},\bm{q}).
\end{equation}
The Hamiltonian of Eq.\eqref{eq:H1phon2mag} describes interactions between non-interacting magnons and non-interacting pho-nons. For later convenience we define the 8$\times$8 $\textit{magnon-}$ $\textit{phonon}$ $\textit{scattering}$ $\textit{matrix}$ (4$\times$4 for the N\'eel and ferromagnetic phases),
\begin{equation}
\label{eq:scatter}
{T} ({\bm{k}},{\bm{q}})=\frac{S}{2}\sqrt {\frac{\hbar }{{2NM{\omega _{{\bm{q}}s}}}}} {{{U} }^\dag }({\bm{k}}){\Lambda ' } ({\bm{k}},{\bm{q}}){U} (\bm{k}-\bm{q}).
\end{equation}

\begin{figure}
    \centering
    \includegraphics[scale=0.47]{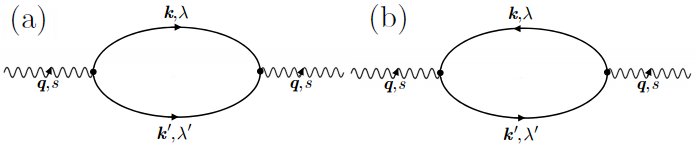}
    \caption{Lowest order magnon-phonon scattering diagrams used for the calculation of the transport relaxation times in the regime in which thermal transport is phonon-dominated. Wavy lines represent phonon propagators whereas straight lines are magnon propagators. Fig.(a) represents C-processes which involve two magnon creations or annihilations, where-as Fig.(b) represents R-processes that involve phonon emission or absorption.}
    \label{fig:phonondiag}
\end{figure}

The magnon-phonon scattering matrix can be partitioned as,
\begin{equation}
\label{eq:tmatrix}
{T} ({\bm{k}},{\bm{q}}) = {\left[ {\begin{array}{*{20}{c}}
  {{{\left[ {{T_{ +  - }}({\bm{k}},{\bm{q}})} \right]}_{4 \times 4}}}&{{{\left[ {{T_{ +  + }}({\bm{k}},{\bm{q}})} \right]}_{4 \times 4}}} \\ 
  {{{\left[ {{T_{ -  - }}({\bm{k}},{\bm{q}})} \right]}_{4 \times 4}}}&{{{\left[ {{T_{ -  + }}({\bm{k}},{\bm{q}})} \right]}_{4 \times 4}}} 
\end{array}} \right]_{8 \times 8}},
\end{equation}
where the submatrices
\[T_{ +  - }({\bm{k}},{\bm{q}})={\left[ {\rm magnon\;creation + annihilation} \right]}_{4 \times 4}, \]
\[T_{ +  + }({\bm{k}},{\bm{q}})={\left[ {\rm two-magnon\;creation} \right]}_{4 \times 4}, \]
\[T_{ -  - }({\bm{k}},{\bm{q}})={\left[ {\rm two-magnon\; annihilation} \right]}_{4 \times 4}, \]
\[T_{ -  + }({\bm{k}},{\bm{q}})={\left[ {\rm magnon\;annihilation + creation } \right]}_{4 \times 4}, \]
are related to the Feynman diagram processes discussed in the next section. 
\begin{figure}
    \centering
    \includegraphics[scale=0.5]{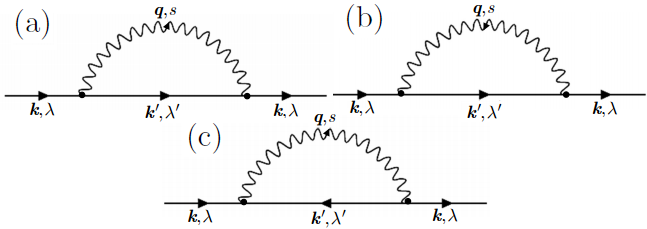}
    \caption{Lowest order magnon-phonon scattering diagrams used for the calculation of the transport relaxation times in the regime in which thermal transport is magnon-dominated. Straight lines represent magnon propagators whereas wavy lines are phonon propagators. Figures (a) and (b) represent R-processes which involve phonon emissions or absorptions. Fig.(c) represents C-processes which involve phonon emission or absorption.}
     \label{fig:magnondiag}
\end{figure}

As seen from Eqs.\eqref{eq:H1phon2mag} and \eqref{eq:tmatrix} one-phonon two-mag-non processes can be classified into two main categories: (a) $\textit{radiation\; processes}$ (denoted as R-processes) and (b) $\textit{conversion\; processes}$ (denoted as C-processes), where the R-processes are described by the submatrices $T_{ +  - }({\bm{k}},{\bm{q}})$ and $T_{ -  + }({\bm{k}},{\bm{q}})$ in which two magnons of the same or different branch, of the same or different valley are involved (one created, one annihilated), whereas the C-processes are described by the submatrices $T_{ +  + }({\bm{k}},{\bm{q}})$ and $T_{ -  - }({\bm{k}},{\bm{q}})$ in which two magnons of the same or different branch, of the same or different valley are either created by a phonon or annihilated into a phonon. Processes described by three boson creation or annihilation operators are not taken into account as they do not conserve energy, which is assumed to be exchanged only between the magnons and the phonons (or they could belong to higher order magnon phonon processes).  Concluding this section, it should be noted that the summations over the phonon and magnon wavevectors in the previous equations extend over the corresponding first Brillouin zones (but in the low temperature regime the main contributions come from the regions around the valleys (minima) of the phonon and the magnon bands), and  that only $\textit{normal processes}$ are taken into account (see Ref.[\onlinecite{Solyom}], section 6.2.4).

\section{\label{sec:mag_pho_relxtimes} Transport relaxation times}
As mentioned previously, in this work we study two distinct thermal transport regimes depending on whether the magnon or the phonon energy scale dominates. In either case, given the matrix elements of the two magnon-one phonon scattering processes, Eq.\eqref{eq:scatter}, one can proceed to calculate the respective transport relaxation times using the Fermi's Golden Rule for each (bare) interaction vertex
\begin{equation}
\label{eq:FRG}
\tau _{I \to F}^{ - 1} =\frac{2\pi}{\hbar}\sum_{F}{\left| {\left\langle F \right|{{\cal \hat H}_{\operatorname{int} }}\left| I \right\rangle } \right|^2}\delta ({E_F} - {E_I}),
\end{equation}  
where $ \left| I \right\rangle $ and $ \left| F \right\rangle $ denote the initial and the final state. 
In the following, we will repeatedly refer to the diagrams of the Figs.~\ref{fig:phonondiag} and~\ref{fig:magnondiag}, denoting a phononic channel as ($\bm{q},s$), and two distinct magnonic channels as ($\bm{k},\lambda$) and ($\bm{k}',\lambda'$). The final state $ \left| F \right\rangle $ for a two-magnon annihilation C-process is
 \begin{align}
 \label{eq:if--}
&{\left| F \right\rangle } = \left| {...,{n_s}( {\bm{q}}) + 1,...} \right\rangle  \otimes \left| {...,{n_\lambda}({\bm{k}}) - 1,} {...,{n_{\lambda'}}(\bm{k}') - 1,...} \right\rangle, 
 \end{align}
 and for a two-magnon creation C-process is
 \begin{align}
 \label{eq:if++}
&{\left| F \right\rangle } = \left| {...,{n_s}( {\bm{q}}) - 1,...} \right\rangle  \otimes \left| {...,{n_\lambda}({\bm{k}})+1 ,} {...,{n_{\lambda'}}(\bm{k}')+1,...} \right\rangle,
 \end{align}
whereas, for a phonon annihilation R-process is 
 \begin{align}
 \label{eq:if-+}
&{\left| F \right\rangle } = \left| {...,{n_s}( {\bm{q}}) - 1,...} \right\rangle  \otimes \left| {...,{n_\lambda}({\bm{k}})-1 ,} {...,{n_{\lambda'}}(\bm{k}')+1,...} \right\rangle, 
 \end{align}
 and for a phonon creation R-process is
 \begin{align}
 \label{eq:if+-}
&{\left| F \right\rangle } = \left| {...,{n_s}( {\bm{q}}) + 1,...} \right\rangle  \otimes \left| {...,{n_\lambda}({\bm{k}})-1 ,} {...,{n_{\lambda'}}(\bm{k}')+1,...} \right\rangle, 
 \end{align} 
 where $\lambda$ and $\lambda'$ denote the same or different magnon ba-nds/branches, $s$ represents any of the two-dimensional acoustic phonons, and finally, momentum conservation (not momentum equivalence as in the $\textit{umklapp processes}$) is applied to each interaction vertex. 
 
In the rest of this section, the transport relaxation times for phonons and magnons are calculated, and are afterwards used in the calculation of the diagonal components of the phonon and magnon conductivity tensor respectively. In this study, we focus only on two relaxation mechanisms appearing during the thermal transport: the magnon-phonon scattering mechanism and the always existent boundary scattering (for either the phonons or the magnons). Depending on the relative strength of the characteristic energy scales of the two types of heat carriers, we further distinguish between two limiting thermal transport regimes, the $\textit{phonon dominated}$ and the $\textit{magnon dominated}$, which in turn consist of three subregimes each, the $\textit{diffusive}$, the $\textit{intermediate}$, and the $\textit{ballistic}$ subregime.

\subsubsection{Transport relaxation times for magnon-dominated thermal transport }
In the case in which the magnon characteristic energy dominates, phonons play the role of a bath, and given the assumed weak magnon-phonon coupling, the problem translates into a problem of a system weakly interacting with a bath. The lowest order non-equivalent Feynmann diagrams to be used for the calculation of the transport relaxation times are those appearing in Fig.~\ref{fig:magnondiag}, and focusing on the magnonic channel ($\bm{k},\lambda$), their total contribution is ($s=1$, since as discussed in the subsection~\ref{sssec:magmagdif} below, only the longitudinal acoustic phonon is of interest)

\begin{align}
\label{eq:magreltimes}
&\left. \frac{1}{\tau_{\lambda}(\bm{k})} \right|_{mp}=\frac{2\pi}{\hbar}\sum_{\bm{q},\lambda'} \bigg\{ 
\left|{T_{-+}^{\lambda\lambda'}({\bm{-k,-q}})}\right|^2 \times \nonumber \\
&\times(n_{\bm{q}}+n_{\bm{k-q},\lambda'}+1)\delta(\epsilon_{\bm{k-q},\lambda'}+\hbar\omega_{\bm{q}}-\epsilon_{\bm{k},\lambda}) + \nonumber \\ 
&\left|{T_{-+}^{\lambda\lambda'}({\bm{-k,q}})}\right|^2 (n_{\bm{q}}-n_{\bm{k+q},\lambda'}) \delta(\epsilon_{\bm{k+q},\lambda'}+\hbar\omega_{\bm{q}}-\epsilon_{\bm{k},\lambda})+ \nonumber \\
&\left|{T_{--}^{\lambda\lambda'}({\bm{k,-q}})}\right|^2 (n_{\bm{q-k},\lambda'}-n_{\bm{q}}) \delta(\hbar\omega_{\bm{q}}-\epsilon_{\bm{q-k},\lambda'}-\epsilon_{\bm{k},\lambda}) \bigg\}
\end{align}
where the first term on the right hand side (RHS) of Eq.\eqref{eq:magreltimes} corresponds to the Feynmann diagram of the Fig.~\ref{fig:magnondiag}(a), the second term to the Feynmann diagram of the Fig.~\ref{fig:magnondiag}(b), and the last term to the Feynmann diagram of the Fig.~\ref{fig:magnondiag}(c), and for the magnon-phonon scattering matrix elements we used the convention that follows Eq.\eqref{eq:tmatrix}. Notice that the above result is directly related to the collision integral of the semiclassical Boltzmann transport theory as applied to the system of the magnons within the relaxation time approximation (see Eq.\eqref{eq:relaxtime}).

The calculation of the RHS of Eq.\eqref{eq:magreltimes} proceeds by turning the summation over the phonon wavevectors into an integral using the well-known formula ($A$ stands for the area)
\begin{equation*}
\frac{1}{A}\sum_{\bm{q}}F(\bm{q})=\int\frac{d^2\bm{q}}{(2\pi)^2}F(\bm{q}).
\end{equation*}
It should be noticed though, that the highly anisotropic nature of the magnon band structure (as opposed to the phonon band structure) precludes the analytical solution of the energy constraints imposed by the presence of the Dirac $\delta$ functions in Eq.\eqref{eq:magreltimes}, and one can proceed with the calculation by taking advantage of the $\delta$ function to reduce the dimensionality of the integral by one, by employing the well-known result that 
$\int_{V}{f(\bm{r})\delta[g(\bm{r})]d\bm{r}}=\int_{S}{\frac{f(\bm{r})}{\mid \nabla g(\bm{r})\mid}d\sigma}$, where $S$ is the $(n-1)$-dim surface inside the $n$-dim volume $V$, defined by the constraint $g(\bm{r})=0$, under the condition that $\nabla{g(\bm{r})}\neq 0$. This way, the aforementioned two-dimensional integrals turn into one-dimensional integrals over the lines that satisfy the energy constraints imposed by the respective Dirac $\delta$ functions.  These calculations require a numerical treatment, since neither the Bogoliubov-Valatin transformation nor the energy constraints admit an analytical solution. For more details the reader is referred to the Appendix \ref{sec:appendixc}.

In Eq.\eqref{eq:magreltimes} it was implicitly assumed that the different scattering events, represented by the non-equivalent Feynmann diagrams of Fig.~\ref{fig:magnondiag}, proceed independently. Including further the effect of the boundary scattering of the magnons and assuming that the magnon-phonon scattering processes proceed independently of the boundary scattering, the total probability of scattering for the magnonic channel $(\bm{k},\lambda)$ obeys the following Matthiessen's rule:\cite{AM_book}

\begin{equation}
\label{eq:magtotalreltimes}
\frac{1}{\tau_\lambda(\bm{k})}= \left. \frac{1}{\tau_{\lambda}(\bm{k})} \right|_{mp} +\left. \frac{1}{\tau_{\lambda}(\bm{k})} \right|_{b},
\end{equation} 
where the boundary scattering transport relaxation time (for the magnons) was defined as  $\left. \frac{1}{\tau_{\lambda}(\bm{k})} \right|_{b}=\frac{|\vec{\upsilon}_{\lambda}(\bm k)|}{L}$, where $L$ is the length of the crystal and $\vec{\upsilon}_{\lambda}(\bm {k})$ the group velocity of the $(\bm{k},\lambda)$ magnonic channel. Notice that $\lambda$ (or $\lambda'$) is $\{1,..,4\}$ for the zig-zag and the stripy phase, and $\{1,2\}$ for the N\'eel and the ferromagnetic phase.

\subsubsection{Transport relaxation times for phonon-dominated thermal transport}
In the case in which the phonon characteristic energy dominates, magnons play the role of a bath, and we again have a problem of a system weakly interacting with a bath. The lowest order non-equivalent Feynmann diagrams to be used for the calculation of the respective transport relaxation times are those appearing in Fig.~\ref{fig:phonondiag}, and focusing on the phononic channel $\bm{q}$ (no band index is used here since we focus only on the transverse acoustic phonon, i.e. $s=1$, and the justification for focusing on the tranverse acoustic phonon only is given in the subsection~\ref{sssec:phophodif} below), their total contribution is 

\begin{align}
\label{eq:phoreltimes}
&\left. \frac{1}{\tau(\bm{q})} \right|_{mp}=\frac{2\pi}{\hbar}\sum_{\bm{k}}\sum_{\lambda,\lambda'} \bigg\{ \left|{T_{++}^{\lambda\lambda'}({\bm{k,q}})}\right|^2 \times \nonumber \\
&\times(n_{\bm{k},\lambda}+n_{\bm{q-k},\lambda'}+1)\delta(\epsilon_{\bm{k},\lambda}+\epsilon_{\bm{q-k},\lambda'}-\hbar\omega_{\bm{q}}) + \nonumber \\ 
&\left|{T_{-+}^{\lambda\lambda'}({\bm{-k,q}})}\right|^2 (n_{\bm{k},\lambda}-n_{\bm{k+q},\lambda'}) \delta(\epsilon_{\bm{k+q},\lambda'}-\hbar\omega_{\bm{q}}-\epsilon_{\bm{k},\lambda}) \bigg\}
\end{align}
where the first term on the RHS of Eq.\eqref{eq:phoreltimes} corresponds to the Feynmann diagram of the Fig.~\ref{fig:phonondiag}(a), and the second term to the Feynmann diagram of the Fig.~\ref{fig:phonondiag}(b). For the magnon-phonon scattering matrix elements we again used the convention that follows Eq.\eqref{eq:tmatrix}. Notice that the above result is directly related to the collision integral of the semiclassical Boltzmann transport theory as applied to the system of phonons within the relaxation time approximation (see Eq.\eqref{eq:relaxtime}).

The calculation of the RHS of Eq.\eqref{eq:phoreltimes} proceeds as in the previous section, i.e. by turning the summation over the magnon wavevector into a two dimensional integral. In Eq.\eqref{eq:phoreltimes} it was implicitly assumed that the different scattering mechanisms, represented by the non-equivalent Feynmann diagrams of  Fig.~\ref{fig:phonondiag}, proceed independently. Including the effect of the boundary scattering of phonons, and assuming that the magnon-phonon scattering processes proceed independently of the boundary scattering, the total probability of scattering for the phononic channel $\bm{q}$ obeys the following Matthiessen's rule:
\begin{equation}
\label{eq:phototalreltimes}
\frac{1}{\tau(\bm{q})}= \left. \frac{1}{\tau(\bm{q})} \right|_{mp} +\left. \frac{1}{\tau(\bm{q})} \right|_{b},
\end{equation} 
where the boundary scattering transport relaxation time (for the phonons) was defined as  $\left. \frac{1}{\tau(\bm{q})} \right|_{b}=\frac{|\vec{\upsilon}_{s}|}{L}$, where $\vec{\upsilon_{s}}$ denotes the phonon group velocity within the approximation of the Debye model. Notice that $\lambda$ (or $\lambda'$) is $\{1,...,4\}$ for the zig-zag and the stripy phase, and $\{1,2\}$ for the N\'eel and the ferromagnetic phase.

\subsubsection{Computational details of the calculation of the transport relaxation times within different transport subregimes}

The calculation of the transport relaxation times requires, via the magnon-phonon scattering matrix elements, knowledge of the spatial derivatives of the Heisenberg and the Kitaev exchange couplings, denoted as $J'$ and $K'$ respectively. For simplicity the derivatives of the exchange couplings are taken as direction independent, and further they are approximated as \cite{Woods:prb01} $J'\approx \frac{\Delta J}{\alpha} \approx \frac{J}{\alpha}$ and  $K'\approx \frac{\Delta K}{\alpha} \approx \frac{K}{\alpha}$, respectively, where $\alpha$ denotes the interionic distance. Based on those definitions, one can convert the integrals appearing in the $\textit {total transport relaxation times}$ (magnonic or phononic) into dimensionless integrals as below
\begin{equation}
\label{eq:taumpestimate}
\frac{1}{\tau_{mp}}\simeq \frac{S A}{E_D} \frac{1}{N_u\alpha^2}\times 2 \times 10^{12} (secs^{-1}) \times I,
\end{equation} 
where, for a specific material, different magnon-phonon scattering processes are encapsulated in the parameter $I$. $S$ denotes the spin of the local moments, $A$ is the energy scale parameter defined in Eq.\eqref{eq:HK_A}, and further, $SA$ defines an appropriate magnonic energy scale dictated by the interaction term of Eq.\eqref{eq:interactionmatrix}, $E_D\equiv \hbar \upsilon_D q_D =\hbar \upsilon_D 2 \pi /\alpha\sqrt{3}$ is the Debye energy scale, $N_u$ the number of nucleons of the ions that form the honeycomb lattice, 
$\alpha$ the interionic distance in Angstroms, and finally $I$ is the dimensionless form of the $\textit {total transport relaxation time}$ (magnonic or phononic).


The relative strength of the magnon-phonon and the boundary scattering for the case of the magnon-dominated thermal transport, can also be written in terms of the dimensionless parameter $I$ mentioned above, as 
\begin{equation}
\label{eq:ratiompmagbestimate}
\left. \frac{\tau_{b}}{\tau_{mp}} \right|_{mag}\simeq c_{mag} \times \frac{1}{\upsilon_{mag}} \times I,
\end{equation}
where $c_{mag}\equiv 55 \times \frac{1}{\Theta_D(K)}\times \frac{L}{\alpha} \times
\frac{1}{N_u\alpha^2}$ , $L$ is the length of the crystal in the direction of the applied temperature gradient, and $\Theta_D$ the Debye temperature in Kelvins. In addition, $\vec{\upsilon}_{mag}(\bm{k})$ is the dimensionless magnon group velocity which is extracted from the dimensional magnon group velocity $\vec{V}_{mag}(\bm{k})$ as below (i.e. their magnitudes are related as)
\begin{align}
\label{eq:vmagdimensionless}
&|\vec{V}_{mag}(\bm{K})|=|\nabla_{\bm{K}} \Omega(\bm{K})| \nonumber \\  
&=\frac{1}{\hbar}\sqrt{\bigg(\frac{\partial(\hbar \Omega(\bm{K}))}{\partial K_x}\bigg)^2+\bigg(\frac{\partial(\hbar \Omega(\bm{K}))}{\partial K_y}}\bigg)^2 \nonumber \\ 
&=\frac{SA\alpha \sqrt{3}}{2\pi\hbar}\sqrt{\bigg(\frac{\partial(\hbar \omega(\bm{k}))}{\partial k_x}\bigg)^2+\bigg(\frac{\partial(\hbar \omega(\bm{k}))}{\partial k_y}}\bigg)^2 \nonumber \\ 
&\equiv \frac{SA\alpha \sqrt{3}}{2\pi\hbar} |\vec{\upsilon}_{mag}(\bm{k})|,
\end{align}
where $\hbar \Omega(\bm{K})$ denotes the dimensional magnon energy and $\hbar \omega(\bm{k})$ the dimensionless magnon energy, the two related as $\hbar \Omega(\bm{K})=SA \times \hbar \omega (\bm{k})$. $\bm{K}$ denotes the dimensional magnon wavevector and $\bm{k}$ the dimensionless one, the two related as $\bm{K}=\frac{2\pi}{\alpha \sqrt{3}}\bm{k}$ ($ \alpha$ is the interionic distance on the honeycomb lattice). 

On the other hand, the relative strength of the magn-on-phonon and the boundary scattering for the case of the phonon-dominated thermal transport can be written in terms of the dimensionless parameter $I$ mentioned above as 
\begin{equation}
\label{eq:ratiompbphoestimate}
\left. \frac{\tau_{b}}{\tau_{mp}} \right|_{pho}\simeq \frac{SA}{E_D} \times c_{mag}\times I,
\end{equation}
where the various the parameters were defined previously. By varying the parameter $c_{mag}$ above, either by using different systems or by changing the dimensions of a particular system, one can tune the relative strength of the magnon-phonon and boundary scattering, and enter the $\textit ballistic$ (boundary scattering dominated), the $\it diffusive$ (magnon-phonon scattering dominated) or the $\textit intermediate$ (competing magnon-phonon and boundary scattering) heat transport subregime.

\section{Calculation of the diagonal components of the thermal conductivity tensor}
\label{sec:thermal_conductivity}

In the previous sections we introduced the low energy magnetic degrees of freedom via the  Hamiltonian of Eq.\eqref{eq:HKham}, the low energy ionic degrees of freedom via the Hamiltonian of Eq.\eqref{eq:Hphonon}, and the magnon-phonon coupling via the Eqs. \eqref{eq:jexpand} and \eqref{eq:kexpand}. In the next step, the magnon spectra of the various ordered phases were computed within the linear spin wave approximation leading to Eq.\eqref{eq:spinwave}, and those spectra were then used as inputs for the lowest order magnon-phonon scattering processes encompassed in the Hamiltonian of Eq.\eqref{eq:H1phon2mag}. The last information was then used to determine via the Fermi's Golden rule the momentum-dependent total transport relaxation times given by Eqs.\eqref{eq:magtotalreltimes} and \eqref{eq:phototalreltimes}, and in the final step   
all those results are patched together to compute the diagonal components of the thermal conductivity tensor for each one of the ordered magnetic states, by using the semiclassical Boltzmann transport theory.


As elaborated in the Appendix~\ref{sec:level5} [see Eq.\eqref{eq:thermalconductivity1}], the$\textit{ thermal conductivity tensor per unit area}$, for heat transport dominated by one type of carriers,  is given by
\begin{equation}
\label{eq:thrmlcndexprsn}
\kappa_{\alpha \beta}= 
\sum\limits_\Lambda {\int\limits_{} {\frac{{d^2{\bm{K}}}}{{{{(2\pi )}^2}}}} }    \hbar {\Omega_\Lambda}({\bm{K}}) v_{\Lambda}^{\alpha}({\bm{K}})   v_{\Lambda}^{\beta}({\bm{K}})  {\tau_\Lambda}({\bm{K}}) \frac{{\partial n_\Lambda^0({\bm{K}})}}{{\partial T}},
\end{equation}
where $\Lambda$ denotes the band index, $\bm{K}$ the wavevector of the quasiparticle, $\hbar {\Omega _\Lambda}({\bm{K}})$ the unrenormalized (in this study) quasiparticle energy, $v_\Lambda^\alpha({\bm{K}})$ the $\alpha$-th component of the quasiparticle group velocity, and $\tau_\Lambda(\bm{K})$ refers to the total transport relaxation time of the dominant carrier. $\frac{{\partial n_\Lambda^0({\bm{K}})}}{{\partial T}}$ denotes the temperature gradient of the equilibrium Bose-Einstein distribution function of the dominant heat carrier. In this work, only the diagonal components of the thermal conductivity tensor per unit area $\kappa_{xx}$ and $\kappa_{yy}$ (with the spatial directions $x$ and $y$ defined as in Fig.\ref{fig:zigzagcell}, Appendix~\ref{sec:magnon_spectrum}) are studied, which are not generally expected to be equal to each other due to the strong spatial anisotropy of the low energy magnetic degrees of freedom, which is further imprinted on the spectra of the low energy magnetic excitations of the various ordered phases.

In the following sections, first the effect of the pure boundary scattering on the heat transport is examined by taking both the magnon and the phonon heat carriers into account. Afterwards, the effect of the magnon-phonon scattering mechanism (to lowest order the two magnon one phonon scattering mechanism) is taken into account on top of the pure boundary scattering, but in order to simplify the whole treatment this work focuses on two simple limiting cases, the phonon dominated and the magnon dominated regime, in which only one type of heat carriers dominates the thermal conductivity. Within the two aforementioned regimes, both scattering mechanisms (boundary and magnon-phonon) are examined for the dominant heat carrier. 

\subsection{Boundary scattering dominated/fully ballistic regime}
In this section, the $\textit{ballistic}$ behavior of the diagonal components  $\kappa_{xx}$ and $\kappa_{yy}$ of the thermal conductivity tensor per unit area is studied, for all the ordered phases of the nearest-neighbor Heisenberg-Kitaev Hamiltonian, versus temperature, for three different relative strengths of the magnon and the phonon characteristic energy scales. Temperature is measured in units of $[T]=\frac{S A}{k_B}$, where $S$ is the spin of the local moment, $k_B$ the Boltzmann constant, and $A$ the magnetic energy scale defined in Eq.\eqref{eq:HK_A}. For the numerical calculations, it is more convenient to convert the ballistic magnon conductivity per unit area to a dimensionless one,  

\begin{eqnarray}
\label{eq:ballmag}
\kappa_{mag}^{ball} =\frac{1}{2\pi}\frac{L}{a}\frac{k_B S A}{\hbar} \times \tilde \kappa_{mag}^{ball},
\end{eqnarray} 
where $\tilde \kappa_{mag}^{ball}$ is the dimensionless ballistic magnon thermal conductivity per unit area and $a=\alpha\sqrt{3}$, where $\alpha$ denotes the interionic distance,
and the same is done for the ballistic phonon conductivity,
\begin{eqnarray}
\label{eq:ballpho}
\kappa_{pho}^{ball}=\left(\frac{E_D}{S A}\right)^3\frac{1}{2\pi}\frac{L}{a}\frac{k_B S A}{\hbar} \times \tilde \kappa_{pho}^{ball},
\end{eqnarray}
where $ \tilde \kappa_{pho}^{ball}$ is the dimensionless ballistic phonon thermal conductivity, and the rest of the notation is known. The $\textit{total ballistic thermal conductivity}$ is
\begin{equation}
\label{eq:balltot}
\kappa_{tot}^{ball}=\kappa_{mag}^{ball}+\kappa_{pho}^{ball}.
\end{equation}
Eqs.\eqref{eq:ballmag}, \eqref{eq:ballpho} and \eqref{eq:balltot} are applied to each of the diagonal components of the conductivity tensor independently, and the results are shown in Figs. \ref{fig:balisticfigxx} and \ref{fig:balisticfigyy} below.

\begin{figure}[htb]
\centering
 \begin{tabular}{@{}cccc@{}}
    \includegraphics[width=.4\textwidth]{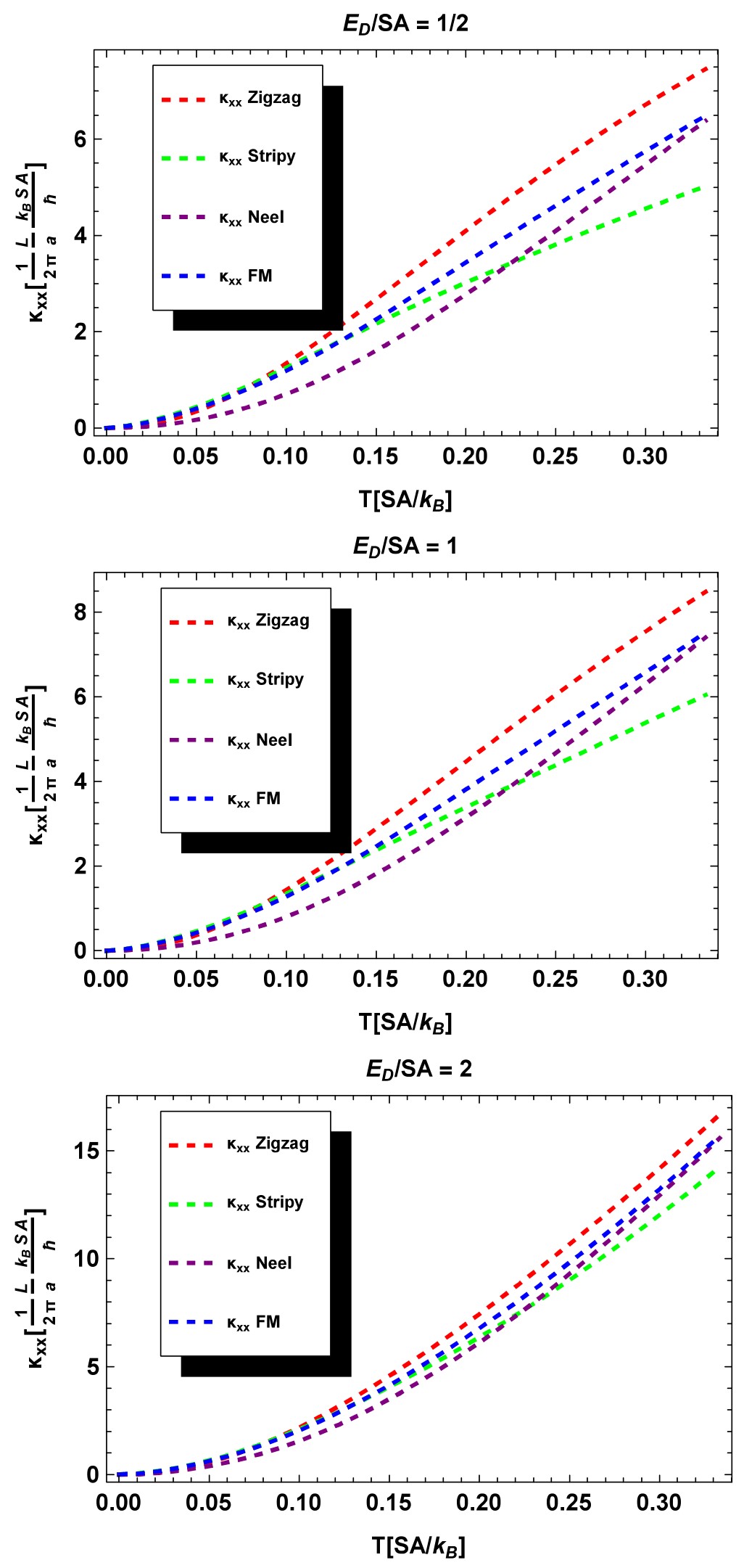} 
 \end{tabular}
  \caption{(Color online) $\kappa_{xx}$ component of the total fully ballistic thermal conductivity per unit area, for each ordered phase (see the legend of each subfigure), for different relative strengths of the Debye energy $E_D$ to the magnon characteristic energy $S A$ (given on top of each subfigure), versus temperature. The temperature region is well below the lowest of the two characteristic energy scales (phononic or magnonic). Notice that the conductivity components are measured in the units given by the prefactor on the RHS of Eq.\eqref{eq:ballmag}. The spatial direction $x$ is defined as in Fig.\ref{fig:zigzagcell}, Appendix~\ref{sec:magnon_spectrum}.}
   \label{fig:balisticfigxx}
\end{figure}

\begin{figure}[htb]
\centering
 \begin{tabular}{@{}cccc@{}}
    \includegraphics[width=.42\textwidth]{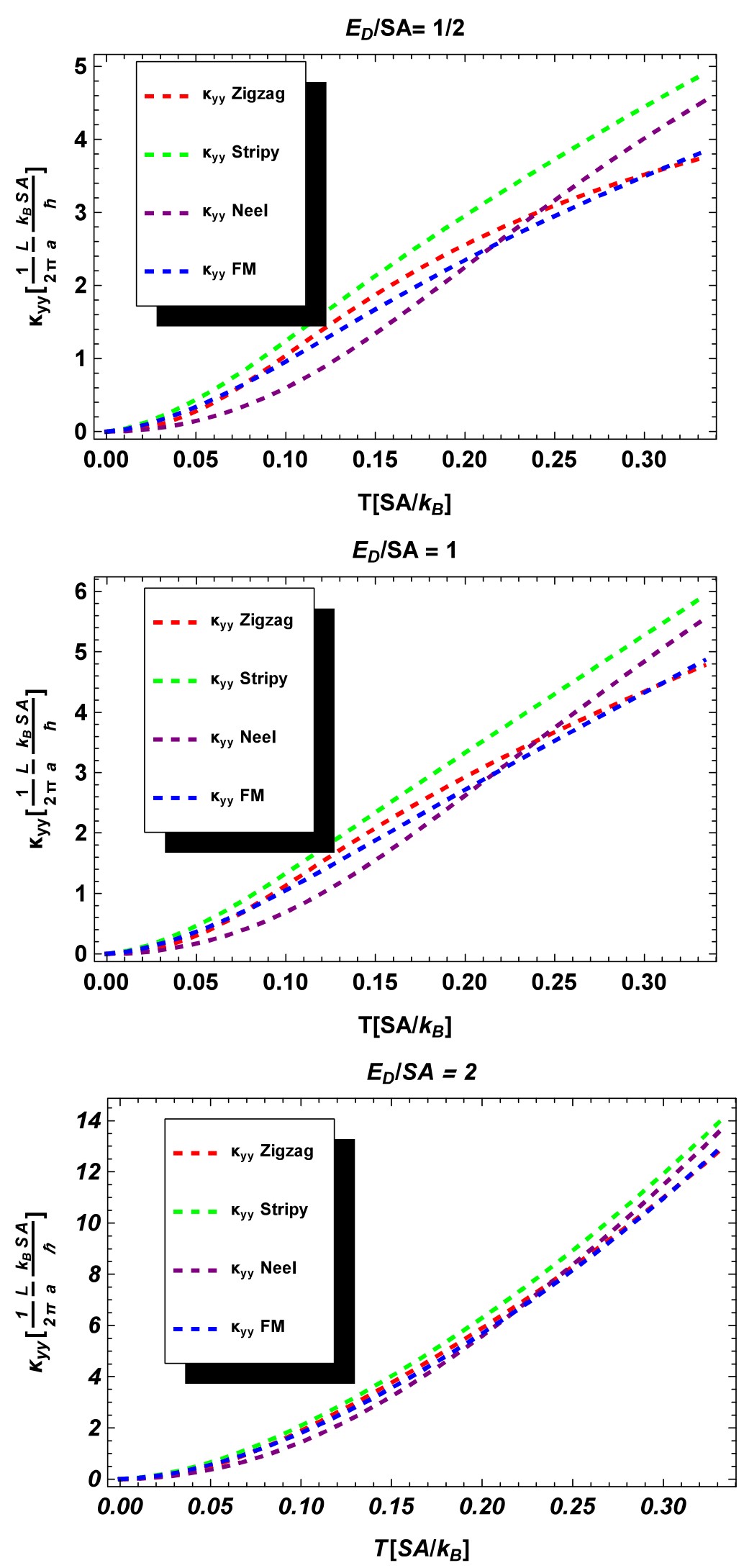} \\
 \end{tabular}
  \caption{(Color online) $\kappa_{yy}$ component of the total ballistic thermal conductivity per unit area, for each ordered phase (see the legend of each subfigure), for different relative strengths of the Debye energy $E_D$ to the magnon characteristic energy $S A$ (given on top of each subfigure), versus temperature. The temperature region is well below the lowest of the two characteristic energy scales (phononic or magnonic). Notice that the conductivity components are measured in the units given by the prefactor on the RHS of Eq.\eqref{eq:ballmag}. The spatial direction $y$ is defined as in Fig.\ref{fig:zigzagcell}, Appendix~\ref{sec:magnon_spectrum}.}
   \label{fig:balisticfigyy}
\end{figure}

In Figs. \ref{fig:balisticfigxx} and \ref{fig:balisticfigyy}, the temperature region was chosen well below the magnetic transition as well as the Debye temperature, since our system of study is assumed to have well-defined low energy magnetic degrees of freedom (given by the Heisenberg-Kitaev Hamiltonian) as well as low energy magnetic excitations. For sufficiently low temperatures the use of the Debye model in the calculation of the phononic thermal conductivity as well as the neglect of higher order processes (magnon-magnon, phonon-phonon or magnon-phonon) are all well justified. Three distinct characteristic energy relative strengths are considered: (i) magnetic energy half the Debye energy, (ii) magnetic energy equal to the Debye energy, and (iii) magnetic energy twice the Debye energy. In the application of Eq.\eqref{eq:thrmlcndexprsn}, the mean free path for the boundary scattering is taken as $\lambda=|\vec{\upsilon}(\bm{K})|\tau(\bm{K})\approx|\vec{\upsilon}(\bm{K})|\frac{L}{|\vec{\upsilon}(\bm{K})|}=L$, i.e. approximately equal to the length of the crystal $L$ (for simplicity we assume a square crystal). In this case, the magnonic and the phononic contribution to the $\textit{total fully ballistic}$ thermal conductivity tensor depends solely on the respective carrier's band structure, via its energy dispersion relation, its group velocity, and its Bose-Einstein occupation factor. Therefore, any differences among the total fully ballistic thermal conductivities directly reflect differences in the carrier band structures, and particularly differences in the magnon band structures, since the phonon band structure is common to all magnetically ordered phases.

Note that for the ferromagnetic and the N\'eel phases there are two magnon bands, while for the zig-zag and the stripy phases there are four bands. In the low temperature region, as far as the magnon contribution is concerned, any gapless magnon bands are more important than any gapped ones, and furthermore, the respective thermal conductivity contribution is dominated by the band structure nearby any magnon valleys. This is because the Bose-Einstein occupation factors decrease rapidly with increasing excitation energies. For a two-dimensional system, there is one longitudinal and one transverse acoustic phonon (notice that in the fully ballistic regime studied in this section both acoustic phonons are taken into account), both assumed to obey a linear isotropic dispersion relation. Particularly, the two acoustic phonons are treated within the Debye model adjusted to a two dimensional system. It is worth noting that the phononic thermal conductivity within the Debye model in the low temperature regime becomes $\propto T^2$ for a 2D system (in contrast to the $T^3$ result for a 3D system). 

Taking into account the previous discussion we now turn our attention to the Figs. \ref{fig:balisticfigxx} and \ref{fig:balisticfigyy}. The bottom diagram of each figure represents phonon dominated total thermal conductivity results. The curves for different phases tend to converge to each other as a result of the common phonon band structure, tend to follow a parabolic dependence on the temperature as a result of the Debye model applied to 2D systems, and tend to become more isotropic due to the smaller difference between the values of the $\kappa_{xx}$ and $\kappa_{yy}$ components. The top diagram of each figure represents magnon dominated total thermal conductivity results that differ appreciably from each other over different magnetic phases as a result of the very different magnon band structures of the various phases, deviate significantly from the parabolic dependence on the temperature which is characteristic of the in-plane acoustic phonons, and tend to become significantly anisotropic due to the greater difference between the values of the $\kappa_{xx}$ and $\kappa_{yy}$ components. The intermediate diagram of each figure corresponds to an $\textit{intermediate fully ballistic subregime}$ which, as far as the boundary scattering dominated heat transport is concerned, is characterized by comparable magnonic and phononic contributions.

Another point to notice is that, away from the phonon dominated regime, even though as $T\to0$ all conductivities go to zero, there is a temperature window from $0.05 \frac{SA}{k_B}$ up to about $0.11 \frac{SA}{k_B}$ within which the N\'eel total conductivity is markedly lower than the total conductivities of the other three magnetic phases which have higher but nearby values. In addition, the total conductivity of the N\'eel phase (either component) seems to saturate slower than all the other conductivities within the examined temperature window. This can be traced back to the structure of the lowest magnon bands of the different ordered states. As will be detailed below, the particular feature of the N\'eel state is that it has a nearly isotropic magnon band whose minimum is located at the center ($\Gamma$ point) of its corresponding 1BZ. However, let us first introduce some useful terminology that will be employed for the structural description of the various magnon bands.

In all the following analysis (and sections) we will use the term $\textit{stiff anisotropy}$ to refer to the gapless magnon bands which approach zero energies with non-zero group velocities (as also happens with the acoustic phonons) and the term $\textit{soft anisotropy}$  to refer to the gapless mag-non bands which approach zero energies with zero group velocities.

For the zig-zag phase, all the four magnon bands are important at low temperatures (since all of them have magnon valleys), and further, half of them are strongly anisotropic whereas the rest half are nearly isotropic. In addition, all four bands have stiff anisotropy and magnon valleys far from the center of the corresponding 1BZ. For the stripy and the ferromagnetic phase on the other hand, only half of their bands are important at low temperatures. In either case the bands are softly anisotropic, and further, the magnon valleys of the stripy phase are away from the center of the corresponding 1BZ, whereas the ferromagnetic phase has its magnon valley at the center of the corresponding 1BZ. Finally, for the N\'eel phase, both of its bands are important at low temperatures, and further, both magnon bands have their minima at the center ($\Gamma$ point) of the corresponding 1BZ, with the one band being stiffly isotropic and the other being stiffly anisotropic around the corresponding valleys. 

As will be seen in the following discussion, the presence of a stiffly (nearly) isotropic band with a valley (minimum) at the center of the corresponding 1BZ ($\Gamma$ point) has some special properties. 
The lowest magnon bands of the ferromagnetic and the stripy phases are $\textit{softly anisotropic}$ which implies that at very low temperatures there can be many more excited magnon quasiparticles compared to the phases which are $\textit{stiffly anisotropic}$. The zig-zag phase is partially $\textit{softly anisotropic}$ and partially $\textit{stiffly anisotropic}$, but all of its magnon bands are gapless which implies more spin wave valleys. As a result, there can again be many excited magnon quasiparticles. 

The N\'eel phase, which is a $\textit{stiffly anisotropic}$ phase with half the bands and fewer valleys ($1/4$) compared to the zig-zag phase, doesn't have any of the aforementioned leeway to increase the population of its low energy magnon quasiparticles (remember that a significant contribution at low temperatures comes from the nearly isotropic valley at the center of the corresponding 1BZ), and this in turn delays the corresponding saturation of its total thermal conductivity (even if it is magnon-dominated). 

Now, given the previous analysis, it seems that the deviation from isotropy (in the magnon bands of interest) leads to a faster saturation of the total conductivity, other than an induced difference between the values of the two diagonal components of the thermal conductivity tensor. It can further be seen from the top and the middle panel of Figs. \ref{fig:balisticfigxx} and \ref{fig:balisticfigyy} that as the temperature increases above zero, the $\kappa_{xx}$ component of the stripy phase and the $\kappa_{yy}$ component of the zig-zag phase seem to saturate first.  The reason for this is that for a temperature gradient along the $x$ direction (Fig.\ref{fig:zigzagcell}, Appendix~\ref{sec:magnon_spectrum}), the stripy phase has its softest magnon modes in that direction, whereas for a temperature gradient along the $y$ direction (Fig.\ref{fig:zigzagcell}, Appendix~\ref{sec:magnon_spectrum}), the zig-zag phase has its softest magnon modes along that direction.

Before concluding this section, it is worth mentioning that close inspection of the top and the bottom diagrams of the Figs. \ref{fig:balisticfigxx} and \ref{fig:balisticfigyy} leads to the additional conclusion that the greater heat current (i.e. the greater contribution to the total thermal conductivity) is carried by the heat carriers with the greater characteristic energy scale (and therefore the greater group velocities). As a measure of the validity of the last statement one can use the low temperature behavior predicted by the two dimensional Debye model (its no saturation sign) as well as the degree of isotropy of the total thermal conductivity that are typical of phonon contributions, and check how the resulting conductivity deviates from the aforementioned typical behavior as one moves toward the magnon dominated side of the fully ballistic regime. 

In the following section we will focus on the effect of the weak magnon-phonon scattering on the magnon dominated and the phonon dominated heat transport, using the results of the fully ballistic regime (magnons and phonons included) examined in this section as a reference.

\subsection{Magnon-phonon dominated/diffusive regime}
\label{ssec:magphodif}

\subsubsection{Phonon dominated heat transport} 
\label{sssec:phophodif}

In this section we focus on the effect of weak magnon-phonon scattering on the phonon dominated heat transport, using the results of the fully ballistic regime as a reference. (From now on, by this term we mean the boundary scattering dominated phononic heat transport.)  One reason for this is that all the factors appearing in the formula of the thermal conductivity tensor are the same for both the boundary and the magnon-phonon relaxation processes, except for the corresponding transport relaxation time. Therefore, any deviations of the thermal conductivity results from the respective pure boundary scattering results are attributed to magnon-phonon scattering (since they originate from transport relaxation times that diverge from the boundary scattering ones). However, before proceeding to the results it would be advisable to first discuss some subtle points that were taken into account in our analysis. 

First, the presumed weak magnon-phonon scattering is to a good extent ensured by working at temperatures much lower than the minimum of the Debye and the magnon characteristic temperature, at which the ionic displacements from their equilibrium positions are small (significantly smaller than a typical lattice constant). Given this, provided that phonon induced changes in the bond lengths and bond angles do not lead to any drastic increase of the gradients of the exchange couplings \cite{Kreizel2011} (if they lead to a drastic decrease as happens in various phenomenological models for the distance dependence of the exchange couplings that does not create any problem),  the magnon phonon couplings $g_{mp}^H \propto \vec{u}_{\bm{q}}\cdot \bm{J'}(\bm{R}_{ij})$ and $g_{mp}^K \propto \vec{u}_{\bm{q}}\cdot \bm{K'}(\bm{R}_{ij})$ are always much smaller than the exchange couplings $\bm{J}(\bm{R}_{ij})$ and $\bm{K}(\bm{R}_{ij})$, respectively.  Then the lowest order perturbative treatment of the magnon-phonon interaction is well justified. Strictly speaking, the distance dependence of the exchange couplings necessitates sophisticated first principle calculations, but keeping in mind that exchange couplings actually originate from electronic exchange paths mediated by neighboring atomic orbital overlaps, an order of magnitude calculation of the gradients of the exchange couplings is feasible and can give an estimate of the strength of the magnon-phonon coupling (see the lines prior to Eq.\eqref{eq:taumpestimate}).

Secondly, as far as the boundary scattering mechanism is concerned, both types of acoustic phonons (transverse and longitudinal) are taken into account via a Debye model adjusted to 2D systems.   As far as the magnon-phonon scattering mechanism is concerned only the transverse phonon is taken into account for the conductivity calculation. The last approximation is tied to the assumption that the magnon-phonon scattering is more important for the longitudinal rather than the transverse (acoustic) phonon, which implies that heat conduction is predominantly borne by the transverse phonon (since the other phonon is scattered too much to contribute to the conduction of the heat and is therefore neglected).   

Another reasoning for this approximation is related to the fact that the main effect of the long-wavelength transverse acoustic phonons is to slightly change/perturb the equilibrium angles between neighboring bonds, whereas, the long-wavelength longitudinal acoustic phonons can change both the equilibrium angles between neighboring bonds (actually depending on their direction of propagation they can be more or less effective), and more important the lengths of the interatomic bonds. As a result, in all cases in which the exchange couplings are much more sensitive to perturbations in the bond lengths (i.e. the radial interionic distances) than in the bond angles, the assumption of a stronger magnon-phonon coupling for the longitudinal acoustic phonon seems to be well justified. 

Having in mind the previous discussion, it is noted that the transverse acoustic phonon is subject to magnon-phonon scattering via a much weaker magnon-phonon coupling constant than the one assumed for the longitudinal acoustic phonon, and this is taken computationally into account by using a $\textit{reduced coupling constant}$ $\tilde g_{mp}(\bm{k,q}) = g_{mp}(\bm{k,q})/\gamma$, where $g_{mp}(\bm{k,q})$ is the magnon-phonon coupling constant used for the longitudinal acoustic phonons, and $\gamma$ is a reduction factor such that $\gamma\sim 10$. Afterwards, the relative strength of the magnon-phonon and boundary scattering, for the (long-wavelength) transverse acoustic phonon assumes, after partitioning it in a dimensional and a dimensionless part, the form (juxtapose with Eq.\eqref{eq:ratiompbphoestimate})
\begin{equation}
\label{eq:ratiompbphoestimatetransverse}
\left. \frac{\tau_{b}}{\tau_{mp}} \right|_{pho}\simeq \frac{SA}{E_D} \times c_{mag} \times \frac{1}{\gamma^2}  \times I,
\end{equation}
where for convenience we set 
\begin{equation}
\label{eq:cphonon}
c_{pho}\equiv \frac{SA}{E_D} \times c_{mag} \times \frac{1}{\gamma^2}.
\end{equation}

Thirdly, for the heat transport process to be phonon dominated, it is legitimately required that the phonon and the magnon energy scales are sufficiently different from each other, and it turns out computationally that a ratio of $E_D/SA = 7$ between the phonon and the magnon energy scales suffices to render the thermal conductivity phonon dominated (by order of magnitude). Under those conditions, as already elaborated in the previous sections, it is sufficient to focus only on one type of heat carriers (in this case the phonons) for an approximate calculation of the thermal conductivity (because $\textit{only}$ the Boltzmann kinetic equation of the dominant heat carriers is employed for the calculation of the thermal conductivity), treating the much less significant heat carriers as a bath with which the dominant heat carriers can exchange energy quasi-elastically (weak system-bath coupling), as well as momentum. Since the characteristic energy scale ratio $E_D/SA$ was incorporated into the newly-defined parameter $c_{pho}$ of Eq.\eqref{eq:cphonon} [$c_{mag}$ was defined right below Eq. \eqref{eq:ratiompmagbestimate}], the $c_{pho}$ expressed as $c_{pho}=10^l$, $l \in \mathbb{Z}$, is treated as a tunable parameter via which one can computationally access the different phonon dominated subregimes: ballistic, intermediate and diffusive, where now, this subcategorization is based on the competition between the boundary and the magnon-phonon relaxation mechanism during the phonon dominated heat transport process. 

\begin{figure}[htb]
\centering
    \includegraphics[width=.49\textwidth]{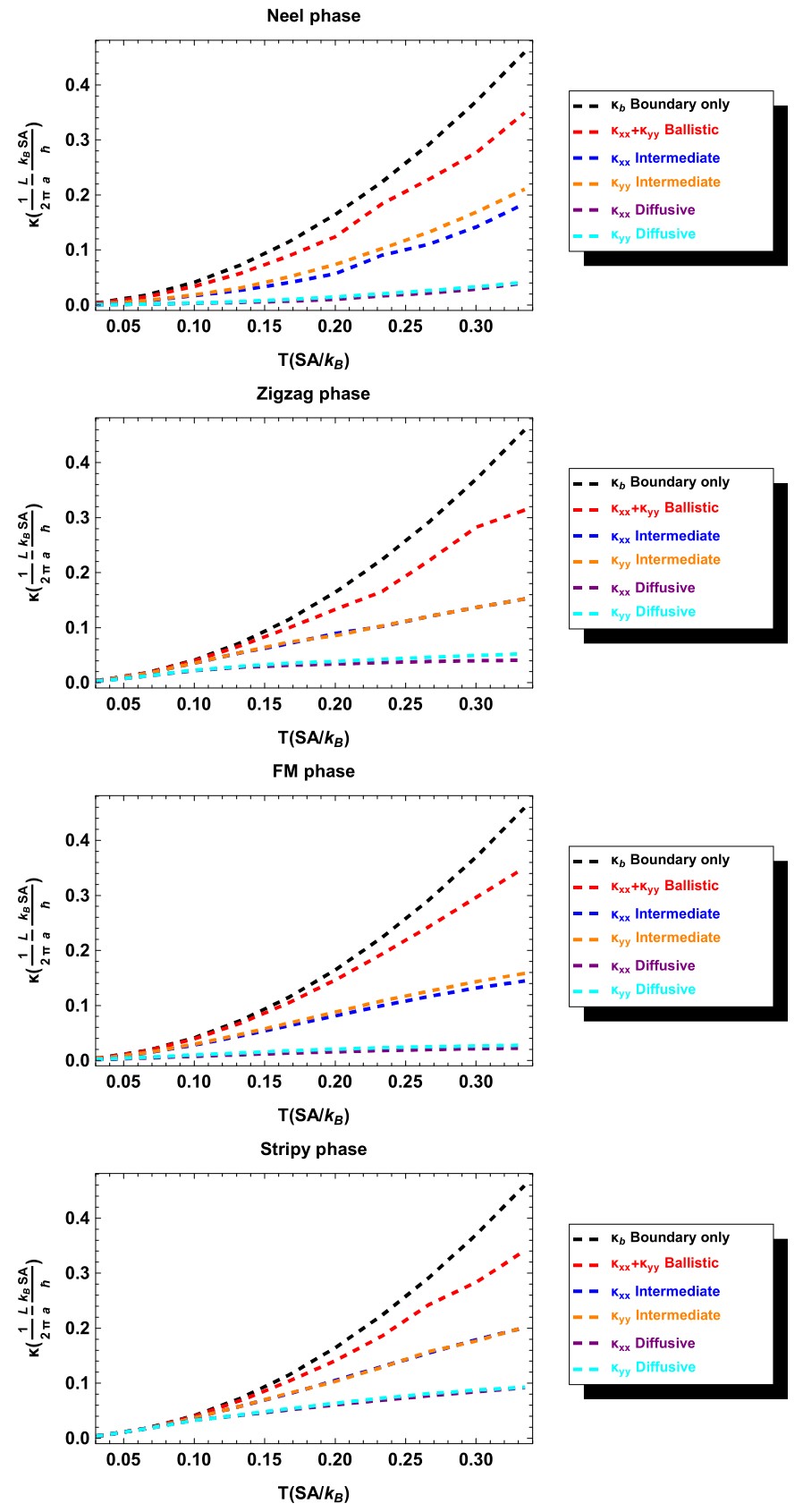} 
  \caption{(Color online) Phonon dominated transport: $\kappa_{xx}$ and $\kappa_{yy}$ component of the phononic thermal conductivity per unit area, for each ordered phase, for three different subregimes: ballistic, intermediate and diffusive (see the legend of each subfigure) as well as pure boundary scattering, versus temperature. The spatial directions $x$ and $y$ are defined as in Fig.\ref{fig:zigzagcell}, Appendix~\ref{sec:magnon_spectrum}.}
   \label{fig:phononfigs1}
\end{figure}

\begin{figure}[htb]
\centering
    \includegraphics[width=.49\textwidth]{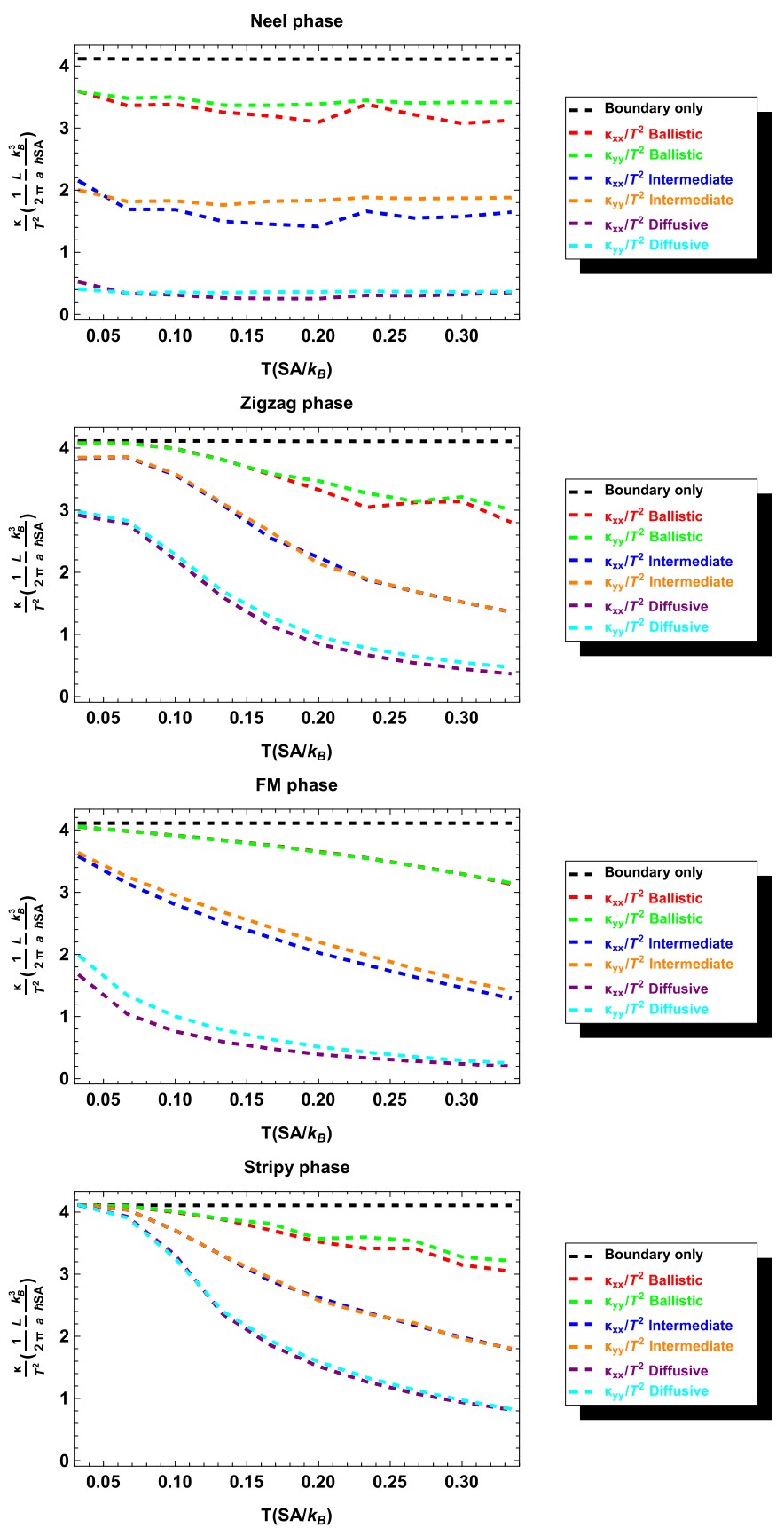} 
  \caption{(Color online) Phonon dominated transport: $\kappa_{xx}/T^2$ and $\kappa_{yy}/T^2$ component of the phononic thermal conductivity per unit area, for each ordered phase, for three different subregimes: ballistic, intermediate and diffusive (see the legend of each subfigure) as well as pure boundary scattering, versus temperature. The spatial directions $x$ and $y$ are defined as in Fig.\ref{fig:zigzagcell}, Appendix~\ref{sec:magnon_spectrum}.}
   \label{fig:phononfigs2}
\end{figure}

Finally, it should be stressed one more time that only sufficiently low temperatures are considered in this work for reasons that were described at various points in the previous analysis (well-defined low energy excitations for the lattice and the magnetic degrees of freedom, weak magnon-phonon coupling, negligible higher order phonon-phonon, magnon-magnon and magnon-phonon processes and so on, are all required to simplify the problem). Particularly, for the phonon dominated heat transport since a lower energy scale is set by the magnons, a rather safe upper limit for the temperature range of interest is set by the magnon characteristic energy, by exploring temperatures smaller than  $T_{max}=\frac{1}{3}\frac{SA}{k_B}$. Respectively, the units of the thermal conductivity are now naturally expressed in terms of the magnon energy scale $SA$ as well. Particularly, in the phonon dominated ballistic (boundary scattering dominated) subregime where the length $L$ of the crystal plays a significant role, the natural units to measure the thermal conductivity are $\frac{1}{2\pi}\frac{L}{a}\frac{k_B SA}{\hbar}$. With all the aforementioned details in mind, let us now turn our attention to the Figs. \ref{fig:phononfigs1} and \ref{fig:phononfigs2} below which show the per unit area components of the phononic conductivity tensor $\kappa_{xx}$ and $\kappa_{yy}$, and
respectively, the same quantities divided by the temperature squared, for each ordered phase, for the three different subregimes mentioned previously (ballistic, intermediate and diffusive, see the legend of each subfigure) as well as for pure boundary scattering, versus temperature.

As already noted above, both acoustic phonons are subject to boundary as well as magnon-phonon scattering.  Since heat conduction is of primary interest, we focus only on the transverse acoustic phonon (the longitudinal one comes in only through boundary scattering, via the 2D Debye model, and its contribution to the heat conduction becomes negligible as one moves away from the purely ballistic deep to the diffusive phonon dominated subregime where it is strongly scattered via the magnon-phonon mechanism). Since the pure boundary scattering for phonons at low temperatures follows exactly the $T^2$ behavior (as a result of the 2D Debye model), Fig.\ref{fig:phononfigs2} actually shows the deviation of the thermal conductivity (due to the transverse long wavelength acoustic phonon) from the $T^2$ low temperature behavior, as one goes from the fully ballistic deep to the diffusive phonon dominated subregime (by tuning the $c_{pho}=10^l$ parameter defined above). In Figs. \ref{fig:phononfigs1} and \ref{fig:phononfigs2} the black curves correspond to boundary scattering dominated phononic heat transport, whereas the red (and the green in Fig.\ref{fig:phononfigs2}) curves correspond to the ballistic subregime, where this term now refers to a situation in which the phononic heat transport is mostly (but not purely) boundary scattering dominated. In the intermediate subregime, as already stressed above, both scattering mechanisms (boundary and magnon-phonon) affect the transverse long wavelength acoustic phonon. It is worth noting that the cross-over from the purely ballistic to the diffusive subregime takes place by gradually decreasing the strength of the boundary scattering (i.e. by increasing the length of the crystal), and along the way the magnon-phonon scattering mechanism is gradually unmasked until it dominates over the boundary scattering mechanism, deep in the diffusive subregime.  

From Figs. \ref{fig:phononfigs1} and \ref{fig:phononfigs2}, it is easily observed that the thermal conductivity is actually isotropic for $\textit{all}$ subregimes, since the $\kappa_{xx}$ and the $\kappa_{yy}$ components almost coincide with each other. This is a consequence of the (intrinsic) isotropic nature of the phonon band structure. Absolute coincidence, upon deviation from the boundary scattering dominated (or fully ballistic) subregime, is not possible  because of the interplay with the magnon bath whose band structure is strongly anisotropic. The qualitative conclusion is that in the phonon dominated regime, no matter how anisotropic the band structure of the magnon bath is, the phonon thermal conductivity succeeds in retaining its isotropic character even deeply in the diffusive subregime of the phonon dominated regime. 

A second striking aspect of the diagrams of Fig.\ref{fig:phononfigs2} (this is hard to be noticed in the diagrams of Fig.\ref{fig:phononfigs1}) is the fact that for all the magnetic phases, except for the stripy phase, the magnon-phonon scattering mechanism seems to start taking effect at fairly low temperatures. This can be seen by the fact that at the low temperature limit used in the calculations ($T=0.05 SA/k_B$) passing from the fully ballistic to the diffusive subregime (i.e. from the top to the bottom of each subfigure), the values of the phononic thermal conductivity components deviate significantly from the values of the corresponding top black curve which conforms to the $T^2$ low temperature behavior, and this of course is indicative of magnon-phonon scatterings at such low temperatures. This last effect, easily seen, is strongest for the N\'eel phase and weakest for the stripy phase. Saying so, one then is naturally led to the following two qualitative results.

The first qualitative result is that within the phonon dominated regime, at very low temperatures, high energy acoustic phonons can sufficiently effectively be scattered by low energy magnons whose band structure has (at least) a pair of $\textit{stiff}$ gapless magnon bands, of sufficiently different stiffness (the more different the stiffnesses the better).  These magnon bands can be isotropic or anisotropic or both (one isotropic, one anisotropic, as happens in the N\'eel phase), but they must both have their minima (their valleys) at the center of the 1BZ (where the acoustic phonon spectra have their minima as well). This conclusion also agrees with the results of Ref.[\onlinecite{Dixon:prb80}] in which, at very low temperatures (liquid helium temperatures), high energy phonons are scattered (though mildly) by low energy magnons whose band structure consists of a pair of stiff magnon bands, of slightly different stiffness, which are isotropic and both have their minima at the center of the 1BZ. 

The second qualitative result is that within the phonon dominated regime, at very low temperatures, high energy acoustic phonons cannot be scattered by low energy magnons whose band structure consists of gapless magnon bands which are soft, and whose minima (valleys) are non-degenerate, far away from the center of the 1BZ as well as far away from each other. This is exactly the case with the stripy phase, which has two low energy gapless magnon bands on the one hand, but which on the other hand are softly anisotropic, have their valleys far from the center of the 1BZ, and all the valleys are located at different points of the $\bm{k}$-space.  As a result, there is only one softly anisotropic band around each valley whose magnons cannot satisfy energy conservation by interacting with the fast moving phonons. 

In conclusion, we mention that as one passes from the fully ballistic deeply to the diffusive subregime, the phononic thermal conductivity keeps decreasing as a result of the stronger and stronger magnon-phonon scattering compared to the boundary scattering (since the magnon-phonon coupling constant is always weak as discussed previously). The last effect is expected within the model we study since
the magnon bands of whichever magnetic phase (even the lower energy bands of the pha-ses which have well separated in energy magnon bands) do not saturate within the temperature window employed in this analysis.

\subsubsection{Magnon dominated heat transport} 
\label{sssec:magmagdif}

In this section we focus on the effect of the weak magnon-phonon scattering on the magnon dominated heat transport, using the results of the fully ballistic subregime (which from now on implies boundary scattering dominated magnonic heat transport) as a reference. Any deviations of the thermal conductivity results from the respective fully ballistic results are attributed to magnon-phonon scattering (since they originate from transport relaxation times that diverge from the purely ballistic ones).  Before proceeding to the results it is helpful to pause and discuss how the arguments presented in the introduction of the previous section are modified for the magnon dominated heat transport that is examined here.

First, the presumed weak magnon-phonon scattering is ensured by working at temperatures much lower than the minimum of the Debye and the magnon characteristic temperature. Second, both for the boundary scattering and the magnon-phonon scattering mechanism $\textit{all}$ magnon bands are taken into account (a Debye-like approximation turns out to be a poor one for the magnons due the highly anisotropic nature and non-linear dispersion of the magnon bands). Third, as far as the magnon-phonon scattering mechanism is concerned only the longitudinal phonon is taken into account for the magnon conductivity calculation, since according to the arguments given in the previous section, any magnon-phonon scattering is predominantly caused by the longitudinal rather than the transverse acoustic phonon. Therefore, the approximation that is adopted is that the diffusive regime of the magnon dominated heat transport originates from the interaction with the longitudinal acoustic phonons. Finally, it should be mentioned that for the heat transport process to be magnon dominated, the phonon and the magnon energy scales must be sufficiently different from each other, and it turns out computationally that a ratio of $SA/E_D = 7$ between the phonon and the magnon energy scales suffices to render the thermal conductivity magnon dominated (by order of magnitude). 

Under those conditions, as was already stressed in the previous sections, it is sufficient to focus only on one type of heat carriers (in this case the magnons) for an approximate calculation of the thermal conductivity (because $\textit{only}$ the Boltzmann kinetic equation of the dominant heat carriers is employed for the calculation of the thermal conductivity).  The less significant heat carriers are then treated as a bath (in this case the phonons) with which the dominant heat carriers can exchange energy quasi-elastically (weak system-bath coupling), as well as momentum. As a reminder, the relative strength of the magnon-phonon and boundary scattering is now given by Eq.\eqref{eq:ratiompmagbestimate}, and the tunable parameter via which one can computationally access the different magnon dominated subregimes  (ballistic, intermediate, diffusive) is the parameter $c_{mag}$ (defined right below Eq.\eqref{eq:ratiompmagbestimate}),  which can more conveniently be expressed as $c_{mag}=10^l$, $l \in \mathbb{Z}$ (to induce order of magnitude changes in the relative strength of the two scattering mechanisms).

\begin{figure}[htb]
\centering
    \includegraphics[width=.48\textwidth]{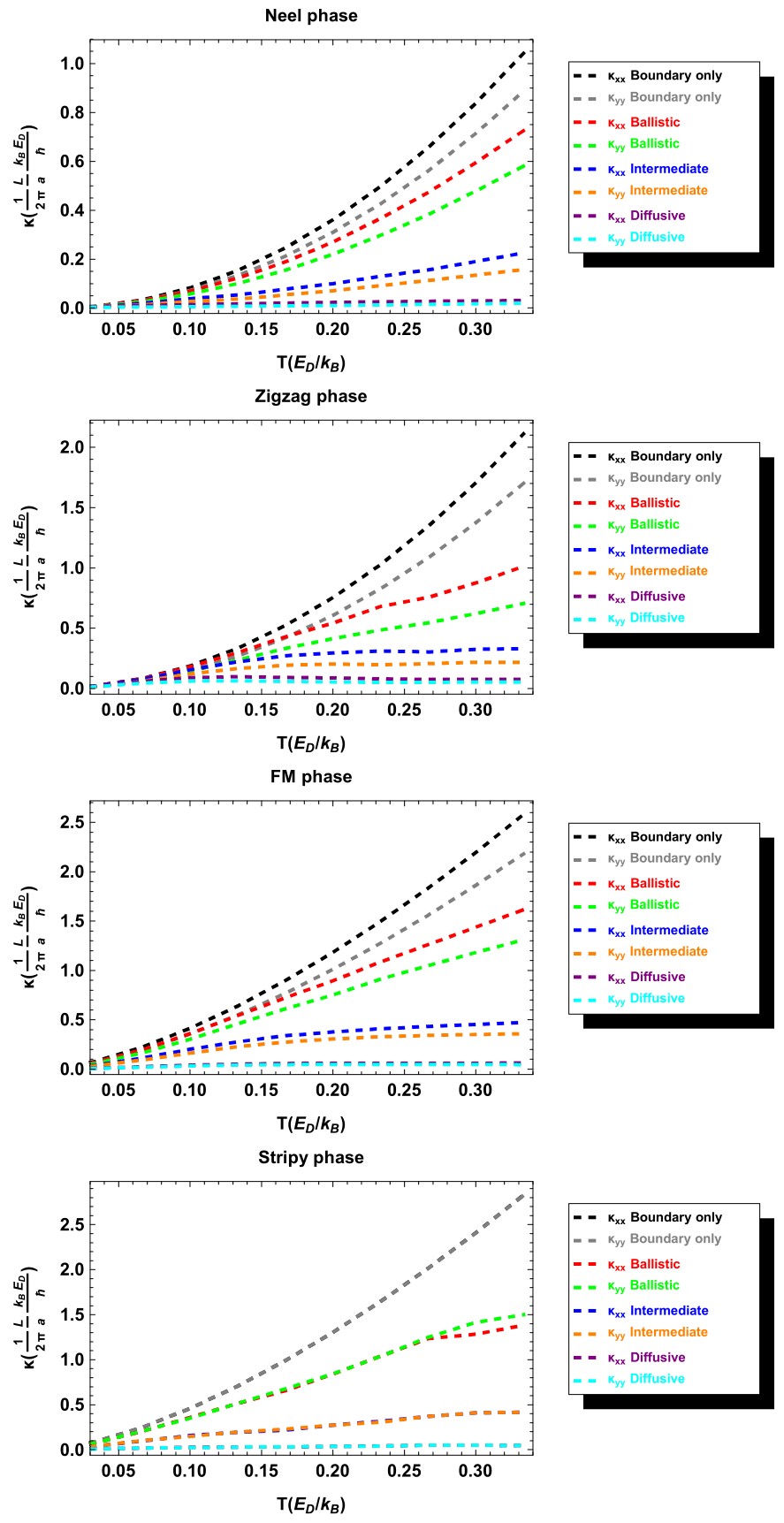} 
  \caption{(Color online) Magnon dominated transport: $\kappa_{xx}$ and $\kappa_{yy}$ component of the magnonic thermal conductivity per unit area, for each ordered phase, for three different subregimes: ballistic, intermediate and diffusive (see the legend of each subfigure) as well as pure boundary scattering, versus temperature. The spatial directions $x$ and $y$ are defined as in Fig.\ref{fig:zigzagcell}, Appendix~\ref{sec:magnon_spectrum}.}
   \label{fig:magnonfig1}
\end{figure}
\begin{figure}[htb]
\centering
    \includegraphics[width=.45\textwidth]{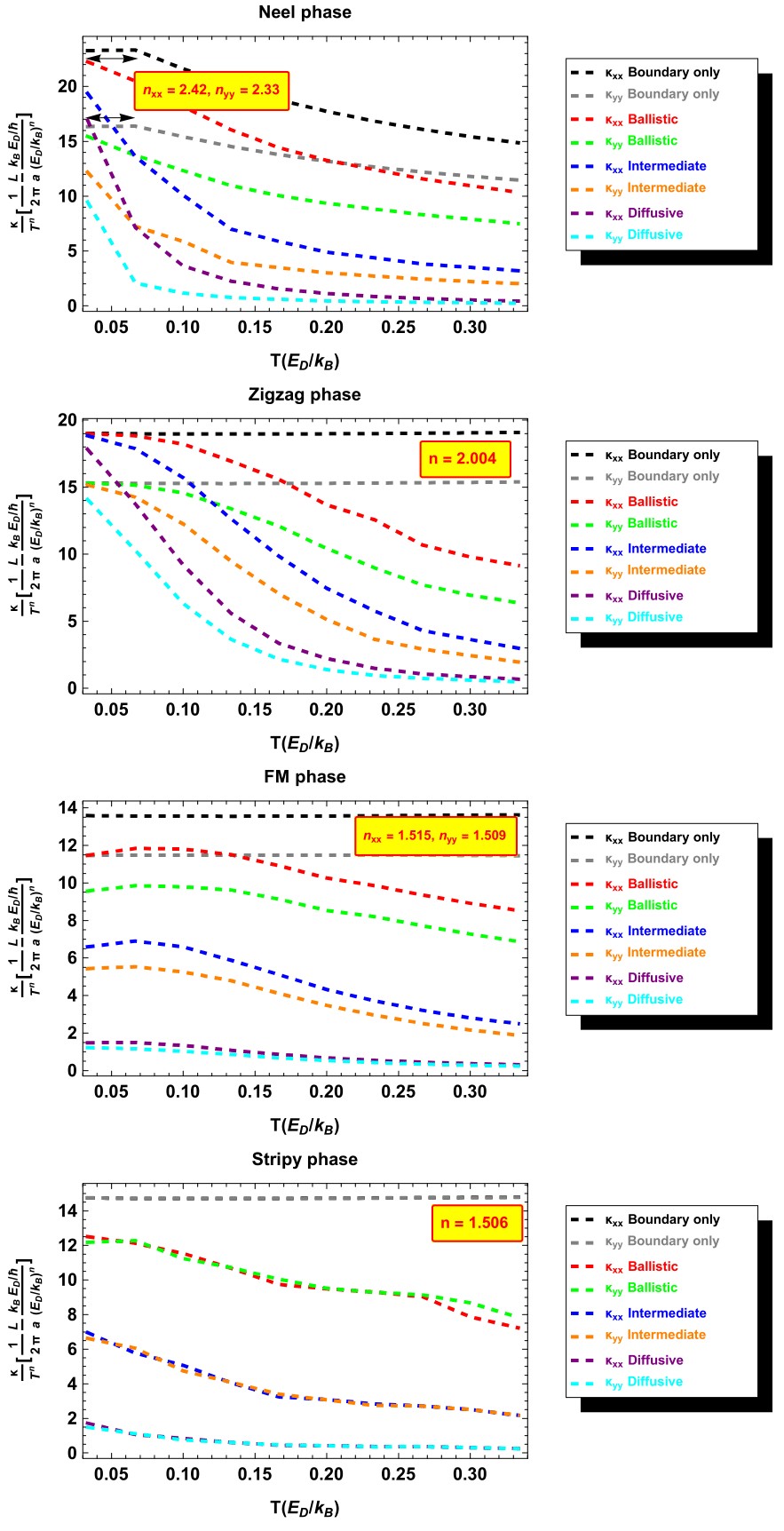} 
  \caption{(Color online) Magnon dominated transport: $\kappa_{xx}/T^n$ and $\kappa_{yy}/T^n$ component of the magnonic thermal conductivity per unit area, for each ordered phase, for three different subregimes: ballistic, intermediate and diffusive (see the legend of each subfigure) as well as pure boundary scattering, versus temperature. The appropriate temperature exponent $n$ that should divide $\kappa_{xx}$ and $\kappa_{yy}$ such that the pure boundary scattering results are represented by horizontal straight lines (at least at low temperatures) is given in the nearby yellow inset. The exponents can slightly vary for the spatial directions $x$ and $y$, as defined in Fig.\ref{fig:zigzagcell}, Appendix~\ref{sec:magnon_spectrum}.}
   \label{fig:magnonfig2}
\end{figure}

We can now turn our attention to the results of Figs. \ref{fig:magnonfig1} and \ref{fig:magnonfig2} which show the behavior of the components of the magnonic thermal conductivity tensor versus temperature, for all ordered phases, for each subregime (ballistic, intermediate, diffusive) as well as pure boundary scattering. In the results of Fig.\ref{fig:magnonfig2} there was an attempt to find a power law for the temperature dependence of the pure boundary scattering mechanism (at least in the low temperature limit of the examined temperature window) so that the respective results lie on a horizontal line (and this is important since by doing so, it is much easier to see the deviations in the results caused by the complementary magnon-phonon scattering mechanism). As can be seen from that figure, the temperature exponent can be slightly different for the $x$ and $y$ directions, as happens for the N\'eel and the FM phase.  Further, for all the magnetic phases except for the N\'eel one, it was possible to find a power of the temperature by which the pure boundary scattering results can be divided so that they all lie along a straight line over the whole examined temperature window (for the N\'eel phase the given exponents cover only the low temperature limit denoted by the horizontal arrows in the top subfigure of Fig.\ref{fig:magnonfig2}). 

Furthermore, from the Figs. \ref{fig:magnonfig1} and \ref{fig:magnonfig2} it is easily seen that the anisotropy of the magnonic conductivity tensor fades away as one moves from the purely ballistic deep to the purely diffusive subregime, and this happens because of the stronger and stronger magnon-phonon scattering from the longitudinal phonons (the deeper we enter the diffusive subregime), or to put it differently, stronger and stronger scattering of the (lower energy) magnons by the (low energy) isotropic longitudinal acoustic phonons gradually washes out any residual anisotropic features of the magnon band structure from the magnonic thermal conductivity. It should further be noticed that the aforementioned effect is stronger for soft low energy magnon bands compared to the analogous effect for stiff low energy magnon bands. Saying so, a qualitative argument that can be given here is that for the softly anisotropic phases (the FM and the stripy phase) the anisotropy of the magnonic conductivity tensor starts diminishing earlier with increasing temperature (i.e. for the aforementioned two phases, the anisotropy of the magnonic conductivity is significantly diminished already at very low temperatures), as opposed to the stiffly anisotropic phases (the zigzag and the N\'eel phase), whose magnon thermal conductivity manages to partially retain the magnon band anisotropies up to higher temperatures. The previous argument is supported by looking at the low temperature side (leftmost side) of each subfigure of Fig.\ref{fig:magnonfig2}, whereby one can see that due to the intense magnon-phonon scattering,  the magnonic conductivity is significantly suppressed compared to its corresponding purely ballistic value. However, the suppression is weaker for the stiffly anisotropic phases (especially for the zigzag phase), where the magnon conductivity is not significantly suppressed from its purely ballistic value unless ones goes to higher temperatures. 

Another feature that one can observe by looking at the subfigures of Fig.\ref{fig:magnonfig1} is that the magnon conductivity of all the magnetic phases seems to saturate within the temperature window employed in this study, except for the N\'eel phase which tends to saturation slower than all the other phases. An explanation for this is that, because the low energy magnon band of the N\'eel phase is stiffly nearly isotropic, with its magnon valley at the center of the 1BZ (where the acoustic phonon bands also have their minima), the strong magnon-phonon scattering mainly affects the lower energy magnons which also have very small wavevectors, whereas the higher energy magnons which are more effective in transporting heat continue to propagate less impeded by the longitudinal acoustic phonons. 

Concluding this section, we emphasize that the magno-nic conductivies of the various magnetic phases differ more markedly from each other closer to the ballistic subregime (or the pure boundary scattering subregime) compared to the diffusive one.  In addition, at very low temperatures (the lower temperature limit of our plots) the boundary scattering mechanism (Fig.\ref{fig:magnonfig2}, see the yellow insets) seems to approximately follow some particular power law, that varies markedly between the stiffly and the softly anisotropic phases (N\'eel and zigzag, and stripy and FM, respectively). A further discrimination between the stripy and the FM phase on the one hand, and the zigzag and the N\'eel phase on the other, deeply within the ballistic subregime, comes from the fact that the values of the two components of the magnon conductivity tensor of the N\'eel and the FM phase follow slightly different power laws (at low temperatures) as opposed to the magnon conductivity components of the other two magnetic phases, which can be described by a common power law.

\section{Conclusions}
\label{sec:conclusions}

In this work we studied the thermal conductivity of electrically insulating local moment models with strong spin-orbit coupling.  As a specific example, we studied the nearest-neighbor Heisenberg-Kitaev model on the honeycomb lattice, whose ground state properties (magnetic orders) are well established.  In particular, for different model parameters, N\'eel, stripy, zig-zag, and ferromagnetic phases are realized.  The richness of the phase diagram originates in the spin-orbit coupling.  For these four magnetic phases, the magnon spectra were initially computed within the linear spin wave approximation. Then, using Fermi's Golden rule in conjunction with the magnon and the phonon spectra, the scattering rates for the lowest order magnon-phonon scattering processes, the two-magnon one-phonon processes, were calculated. Finally, the kinetic Boltzmann equation within the relaxation time approximation was employed for the calculation of the magnonic and the phononic thermal conductivities. The evaluation of the scattering rates was among the most technically challenging aspects of this work, and we had to innovate in order to find an efficient method of computing these rates for the multiple magnon branches. The  procedure we followed and described in this paper can be generalized to  any two-dimensional magnon-phonon system.

Several results and qualitative conclusions for the magnon dominated and the phonon dominated heat transport are contained in Sec.\ref{sec:thermal_conductivity}. We emphasize again that each of the previous regimes is further broken down into three main transport subregimes: the ballistic, the diffusive, and the intermediate subregime. We have also included some discussion of how to estimate which regime may be most relevant to a particular material of a given size. A central result of this analysis is that the effect of the strong spin orbit coupling on the magnetic degrees of freedom, which is to induce anisotropies in the band structures of the low energy magnetic excitations, can most efficiently be probed by measuring the ballistic thermal conductivity of a material whose heat transport is magnon dominated. 

When the phonon energy dominates the magnon energy, the thermal conductivity primarily reflects the spatially isotropic phonon band structure. In this case, the the thermal conductivity tensor remains isotropic, and in the ballistic subregime, at low temperatures, follows a quadratic temperature power law (reminiscent of the 2D Debye model). On the other side, when the magnon energy dominates the phonon energy, the thermal conductivity tensor of the various phases shows significant anisotropic behavior that is strongest within the ballistic subregime. In addition to this, the thermal conductivity of different magnetic phases are found to follow different temperature dependences, even at very low temperatures. 

By carefully analyzing the low temperature dependence and the degree of anisotropy of the thermal conductivity tensor, one may be able to use thermal transport to infer important features of the magnetic order and excitation spectrum that are not easily obtained by other means.  For example, the large neutron absorption cross-section of iridium makes measurements of the magnon spectrum even in bulk iridates difficult. The small signal from resonant inelastic X-ray scattering in a two-dimensional system also makes determination of magnetic order and excitations challenging. Thus, thermal transport may offer a window into the magnetic degrees of freedom where other methods present challenges. On the experimental side, measurements of the thermal conductivity of the SOC-induced Mott insulator $Sr_2IrO_4$ were recently reported \citep{Steckel2016}, which by comparison with the thermal conductivity of the $La_2CuO_4$ antiferromagnet, led to the conclusion that the thermal conductivity of the former is highly suppressed due to strong magnon-phonon coupling, and this effect was then correlated with strong spin-orbit coupling of the iridate.

We hope this work will help stimulate future theoretical and experimental work on thermal transport in insulating local moment systems with strong-spin orbit coupling, since the methodology followed in this work opens a new window to study systems which previously were technically unapproachable. Particularly, magnetically insulating systems that cannot be approached analytically as far as the magnon-phonon interaction problem is concerned, can be numerically approached by the above methodology which relies on the use of a general numerical Bogoliubov transformation for the derivation of the magnon-phonon interaction Hamiltonian and the calculation of magnon-phonon transport relaxation times, even in the presence of anisotropic magnon bands.  These anisotropies are commonplace within the newly discovered field of magnonics, as well as among materials with strong-spin orbit coupling. Such materials may be relevant to applications in spin caloritronics and other spin-based energy, computing, and communications applications. Finally, we note that theoretical estimates of the magnon-phonon relaxation times, that are possible within the above methodology, could be useful to experimentalists who want to know (approximately) the strength of the magnon-phonon relaxation time in their specific systems of study (to the extent that the heat transport is dominated by the mechanisms studied in this work).

\acknowledgements
We gratefully acknowledge discussions with prof. Nicole Benedek and prof. Gregory C. Psaltakis, and funding from ARO grant W911NF-14-1-0579 and NSF DMR-1507621.
\\
\\
\\
\appendix

\begin{figure}
    \centering
    \includegraphics[scale=0.8]{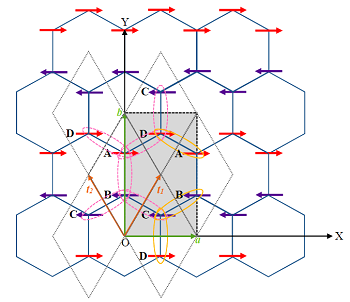}
    \caption{(Color online) Zig-zag magnetic phase: A magnetic unit cell consists of four magnetic moments labelled as A, B, C and D, and is represented by the gray-shaded rectangle shown in the figure. The translation vectors of the periodic magnetic structure are the vectors ${\bm{a}}$ and ${\bm{b}}$. The translation vectors of the chemical periodic structure are the vectors $\bm{t_1}$ and $\bm{t_2}$, and a chemical unit cell is represented by any dashed parallelogram. For the N\'eel and the ferromagnetic states the magnetic unit cell coincides with the chemical unit cell (that is common to all phases).}
    \label{fig:zigzagcell}
\end{figure} 

\section{\label{sec:magnon_spectrum} Linear spin wave theory for the nn Heisenberg-Kitaev Hamiltonian}
In this appendix, we sketch out the derivation of the linear spin wave dispersion relations and the lowest order magnon-phonon scattering amplitudes for the four collinear ordered phases of the Heisenberg-Kitaev model, depending on the relative strength of the Heisenberg and Kitaev couplings through the angle $\varphi$ (see Fig.~\ref{fig:phasediagKH} and Eq.\eqref{eq:HK_A}). The spin wave analysis of the zig-zag and the stripy state requires the use of four magnetic sublattices, labelled as A, B, C, D, and the magnetic unit cell is the rectangular unit cell (gray-shaded rectangle defined by the translation vectors ${\bm{a}}$ and ${\bm{b}}$) shown in Fig.\ref{fig:zigzagcell}. The N\'eel and the ferromagnetic states require only two magnetic sublattices, and the magnetic unit cell coincides with the chemical unit cell of the honeycomb lattice (see the dashed parallelogram whose edges are defined by the translation vectors $\bm{t_1}$ and $\bm{t_2}$ in Fig.\ref{fig:zigzagcell}). Notice that in all the following analysis the spatial gradients of the Heisenberg and the Kitaev exchange couplings are denoted as $\vec {J}^{(1)}$ and $\vec {K}^{(1)}$ respectively.

\subsection{Zig-zag phase}
As already noted, for the zig-zag phase the magnetic unit cell is defined by the gray-shaded rectangle with sides of length $a$ (along the global X-axis) and $b$ (along the global Y-axis), and consists of four magnetic mome-nts A, B, C and D, with A and D pointing along the positive X-axis, and B and C pointing along the nega-tive X-axis. Choosing the positive spin quantization axis along the negative X-axis, at the sites A and D we employ the bosonization given by the Eqs.\eqref{eq:negativeHPoperators1}-\eqref{eq:negativeHPoperators3}, while at the sites B and C we employ the bosonization given by the Eqs.\eqref{eq:positiveHPoperators1}-\eqref{eq:positiveHPoperators3}. Each magnetic site has three nearest neighbors (nn) shown as encircled bonds in Fig.\ref{fig:zigzagcell}. To avoid double counting of the nn interactions, only $\textit {the dashed pink encircled bonds}$ (see Fig.\ref{fig:zigzagcell}) are taken into account. The Kitaev term couples the $z$-spin components along the AB and the CD bond, the $x$-spin components along the upper right AD and the lower left BC bond, and the $y$-spin components along the upper left AD and the lower right BC bond. Using the representation of the $x$- and $y$-spin components in terms of the ladder spin operators to write the total Hamiltonian in terms of the $S_i^{||}$, $S_i^+$ and $S_i^-$ operators, performing the bosonization as elaborated above, and Fourier transforming according to the convention of Eqs.\eqref{Fourieroperators1} and \eqref{Fourieroperators2}, ones finds the classical ground state energy ${\cal H}_{classical}=\frac{NS^2}{2}\left( J - 2K \right) $, and the following $\textit{spin wave mode matrix}$ $M(\bm{k})$ (reference to Eq.\eqref{eq:spinwave} and the notation thereof):
\begin{widetext}
\begin{equation}
{M} ({\bm{k}}) =\left[ {\begin{array}{*{20}{c}}
  A&0&0&{D(\bm{k})}&0&{B(\bm{k})}&0&{C(\bm{k})} \\ 
  0&A&{{D^*(\bm{k})}}&0&{{B^*(\bm{k})}}&0&{{C^*(\bm{k})}}&0 \\ 
  0&{D(\bm{k})}&A&0&0&{C(\bm{k})}&0&{B(\bm{k})} \\ 
  {{D^*(\bm{k})}}&0&0&A&{{C^*(\bm{k})}}&0&{{B^*(\bm{k})}}&0 \\ 
  0&{B(\bm{k})}&0&{C(\bm{k})}&A&0&0&{D(\bm{k})} \\ 
  {{B^*(\bm{k})}}&0&{{C^*(\bm{k})}}&0&0&A&{{D^*(\bm{k})}}&0 \\ 
  0&{C(\bm{k})}&0&{B(\bm{k})}&0&{D(\bm{k})}&A&0 \\ 
  {{C^*(\bm{k})}}&0&{{B^*(\bm{k})}}&0&{{D^*(\bm{k})}}&0&0&A 
\end{array}} \right],
\end{equation}
\end{widetext}
where we defined the following parameters (in this appendix the parameter $A$ appearing in the spin wave mode matrix $M$ should never be confused with the magnetic energy scale defined in Eq.\eqref{eq:HK_A})

\begin{figure}
    \centering
    \includegraphics[scale=0.29]{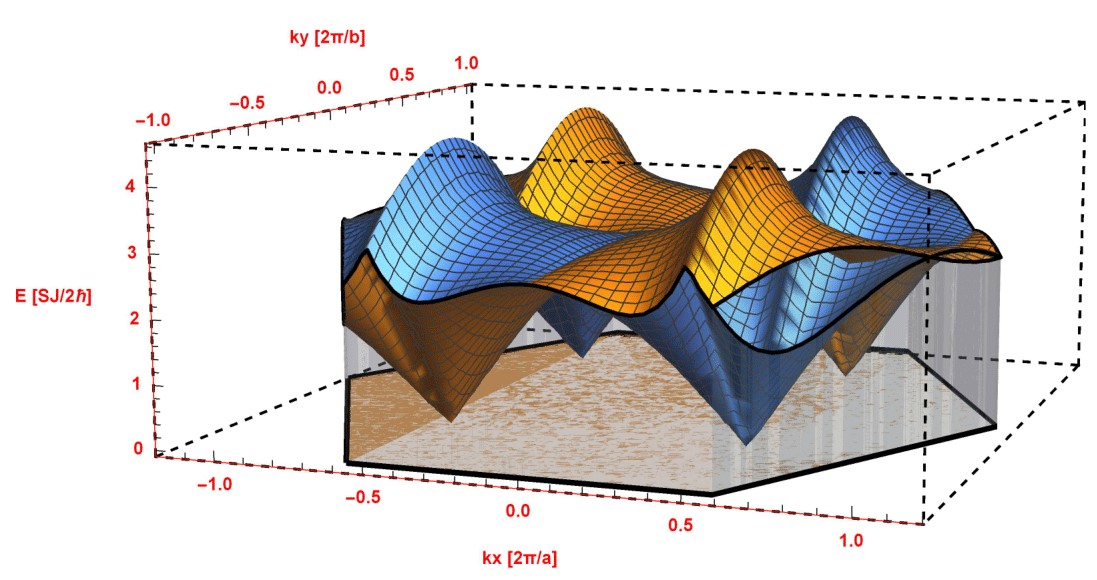}
    \caption{(Color online) Lower spin wave dispersion relations of the zigzag phase, as given by Eqs.\eqref{eq:zigzagdispersions1}. The yellow surface corresponds to $\omega_1(\bm{k})$ and the blue surface to $\omega_2(\bm{k})$. Notice that the magnon wavevector components $k_x$ and $k_y$ are measured in units of $\frac{2\pi}{a}$ and $\frac{2\pi}{b}$ respectively and the spin wave energy is measured in units of $\frac{SJ}{2\hbar}$. The shaded hexagon within the Oxy plane is the first Brillouin zone (1BZ) of the honeycomb lattice. The plot is for $K/J=-2.65$ and $\alpha = 2\pi/3$.}
    \label{fig:zigzagspinwave1}
\end{figure}

 \begin{align*}
  &A = J({{\vec \delta }_1}) - J({{\vec \delta }_2}) - J({{\vec \delta }_3}) + 2K({{\vec \delta }_1})= - J + 2K,   \\
  &B({\bm{k}}) = J({{\vec \delta }_1}){e^{ - i\vec k \cdot {{\vec \delta }_1}}} = J{\eta ^{ - 2},}   \\
  &C({\bm{k}}) = K({{\vec \delta }_3}){e^{ - i\vec k \cdot {{\vec \delta }_3}}} - K({{\vec \delta }_2}){e^{ - i\vec k \cdot {{\vec \delta }_2}}} = 2iK\eta \sin (\pi h),   \\
  &D({\bm{k}}) = \left( {J({{\vec \delta }_3}) + K({{\vec \delta }_3})} \right){e^{ - i\vec k \cdot {{\vec \delta }_3}}} \\
  &+ \left( {J({{\vec \delta }_2}), 
 + K({{\vec \delta }_2})} \right){e^{ - i\vec k \cdot {{\vec \delta }_2}}} 
 = 2(J + K)\eta \cos(\pi h), 
\end{align*} 
in combination with the following definitions

\begin{align*}
&  a = \alpha \sqrt 3,\;\;\;\alpha  = \rm hexagon\;side\; = interionic\;distance,   \\
 & {{\vec \delta }_1} = \frac{1}{3}{\bm{b}} = \frac{1}{3}b{{{\bm{\hat e}}}_Y}, \;\;\;b  = \rm 3\alpha  \\
 & {{\vec \delta }_2} = \frac{1}{2}{\bm{a}} - \frac{1}{6}{\bm{b}} = \frac{1}{2}a{{{\bm{\hat e}}}_X} - \frac{1}{6}b{{{\bm{\hat e}}}_Y},   \\
 & {{\vec \delta }_3} =  - \frac{1}{2}{\bm{a}} - \frac{1}{6}{\bm{b}} =  - \frac{1}{2}a{{{\bm{\hat e}}}_X} - \frac{1}{6}b{{{\bm{\hat e}}}_Y},   \\
 & {\bm{k}} = \left( {h\frac{{2\pi }}{a},\;k\frac{{2\pi }}{b}} \right) = h\frac{{2\pi }}{a}{{{\bm{\hat e}}}_X} + k\frac{{2\pi }}{b}{{{\bm{\hat e}}}_Y},\;\;\;h,k \in \mathbb{Z}   \\
 & \zeta  = {e^{i\pi h}} = {\zeta ^{ - 1}}\;\;\;({\zeta ^2} = 1 = \zeta {\zeta ^{ - 1}}),\;\;\;\eta  = {e^{ik\pi /3 }},\\
 &\bm{t_1}=\frac{1}{2}(\bm{a}+\bm{b}),\;\;\;\bm{t_2}=\frac{1}{2}(\bm{b}-\bm{a}),
\end{align*}
where it is more convenient to measure the components of the magnon wavevector $\bm{k}$ in units of the reciprocal lattice of the magnetic lattice, i.e. in units of $\frac{2\pi}{a}$ and $\frac{2\pi}{b}$ respectively. As far as the parameters $A$, $B(\bm{k})$, $C(\bm{k})$ and $D(\bm{k})$ are concerned, it was assumed that the exchange couplings $J$ and $K$ are bond independent (i.e. the same for each nn bond), as a result of which the bond direction dependence was then dropped.  

\begin{figure}
    \centering
    \includegraphics[scale=0.29]{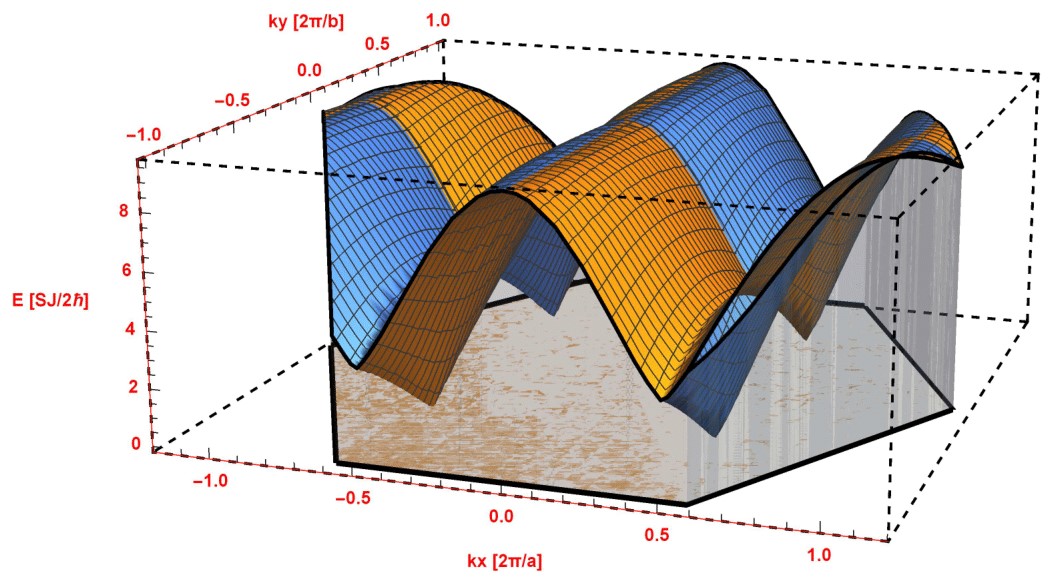}
    \caption{(Color online) Upper spin wave dispersion relations of the zigzag phase, as given by Eqs.\eqref{eq:zigzagdispersions2}. The yellow surface corresponds to $\omega_3(\bm{k})$ and the blue surface to $\omega_4(\bm{k})$. Notice that the magnon wavevector components $k_x$ and $k_y$ are measured in units of $\frac{2\pi}{a}$ and $\frac{2\pi}{b}$ respectively and the spin wave energy is measured in units of $\frac{SJ}{2\hbar}$. The shaded hexagon within the Oxy plane is the 1BZ of the honeycomb lattice. The plot is for $K/J=-2.65$ and $\alpha = 2\pi/3$.}
    \label{fig:zigzagspinwave2}
\end{figure}

Diagonalizing the dynamical matrix $D=I_-M$ as described in Eq.\eqref{eq:dynamicalmatrix} of Sec.\ref{sec:spin_waves}, we obtain the following $\textit{magnon normal modes}$:
\begin{eqnarray}
  \label{eq:zigzagdispersions1}
  \omega _1  = \sqrt {{\Omega _3} - \sqrt {{\Omega _4}} },\;\;\;
  \omega _2  = \sqrt {{\Omega _1} - \sqrt {{\Omega _2}} },\\
  \label{eq:zigzagdispersions2}
  \omega _3  = \sqrt {{\Omega _3} + \sqrt {{\Omega _4}} },\;\;\;
  \omega _4  = \sqrt {{\Omega _1} + \sqrt {{\Omega _2}} },
\end{eqnarray} 
where the following parameters were used
\begin{align*}
 &{\Omega _1} = {A^2} + {\left| D \right|^2} - {\left| {B - C} \right|^2},\\
 & {\Omega _2} = 4{A^2}{\left| D \right|^2} - {\left| {D({B^*} - {C^*}) - {D^*}(B - C)} \right|^2},   \\
 & {\Omega _3} = {A^2} + {\left| D \right|^2} - {\left| {B + C} \right|^2},\\
 &{\Omega _4} = 4{A^2}{\left| D \right|^2} - {\left| {D({B^*} + {C^*}) - {D^*}(B + C)} \right|^2}. 
\end{align*}
The spin wave dispersions of Eq.\eqref{eq:zigzagdispersions1} are plotted in Fig.\ref{fig:zigzagspinwave1} and those of Eq.\eqref{eq:zigzagdispersions2} are plotted in Fig.\ref{fig:zigzagspinwave2}. At low enough temperatures, as far as the magnon-phonon interaction is concerned, only the parts of the spin wave spectra around the $\textit{spin wave valleys}$ are of interest, whose exact $\bm{k}$-space positions are found from the conditions that  $\omega_i=0$, $i={1,2,3,4}$. From Eqs.\eqref{eq:zigzagdispersions1} and \eqref{eq:zigzagdispersions2} it is not hard to see that $\omega_i=0$ implies that either the whole argument of the big (outer) square root is zero, or all the $\Omega_j$ parameters on the respective RHS are simultaneously zero. One can check that the spin wave dispersions of  Eqs.\eqref{eq:zigzagdispersions1} and \eqref{eq:zigzagdispersions2}  have the following symmetry properties
\begin{align*}
 &{\omega_i}(k_x,k_y) = {\omega_i}(-k_x,-k_y)\;\;(\textit{time reversal symmetry}) ,\\
 &{\omega_i}(k_x,-k_y) = {\omega_i}(-k_x,k_y), 
\end{align*}
for $i = 1,2,3,4 $, which can be employed to simplify the calculations. 
As far as the magnon phonon scattering matrix is concerned, the $\Lambda'(\bm{k},\bm{q})$ matrix on the RHS of Eq.\eqref{eq:scatter} has the following form 
\begin{widetext}
\begin{equation}
{\Lambda' } ({\bm{k}},{\bm{q}}) =\left[ {\begin{array}{*{20}{c}}
  {A'(\bm{q})}&0&0&{D'(\bm{k},\bm{q})}&0&{B'(\bm{k},\bm{q})}&0&{C'(\bm{k},\bm{q})} \\ 
  0&{A'(\bm{q})}&{D'(\bm{-k},\bm{q})}&0&{B'(\bm{-k},\bm{q})}&0&{C'(\bm{-k},\bm{q})}&0 \\ 
  0&{D'(\bm{k},\bm{q})}&{A'(\bm{q})}&0&0&{C'(\bm{k},\bm{q})}&0&{B'(\bm{k},\bm{q})} \\ 
  {D'(\bm{-k},\bm{q})}&0&0&{A'(\bm{q})}&{C'(\bm{-k},\bm{q})}&0&{B'(\bm{-k},\bm{q})}&0 \\ 
  0&{B'(\bm{k},\bm{q})}&0&{C'(\bm{k},\bm{q})}&{A'(\bm{q})}&0&0&{D'(\bm{k},\bm{q})} \\ 
  {B'(\bm{-k},\bm{q})}&0&{C'(\bm{-k},\bm{q})}&0&0&{A'(\bm{q})}&{D'(\bm{-k},\bm{q})}&0 \\ 
  0&{C'(\bm{k},\bm{q})}&0&{B'(\bm{k},\bm{q})}&0&{D'(\bm{k},\bm{q})}&{A'(\bm{q})}&0 \\ 
  {C'(\bm{-k},\bm{q})}&0&{B'(\bm{-k},\bm{q})}&0&{D'(\bm{-k},\bm{q})}&0&0&{A'(\bm{q})} 
\end{array}} \right],
\end{equation}
\end{widetext}
and further, for $\textit{long-wavelength acoustic phonons}$ it is 
\begin{align*}
&A'({\bm{q}})=  i \Big[\left( {{{\hat e}_{\bm{q}s}} \cdot {{\vec J}^{(1)}}({{\vec \delta }_1})} \right)\left( {\vec q \cdot {{\vec \delta }_1}} \right)
 -\left( {{{\hat e}_{\bm{q}s}} \cdot {{\vec J}^{(1)}}({{\vec \delta }_2})} \right)\left( {\vec q \cdot {{\vec \delta }_2}} \right)\\
 &-\left( {{{\hat e}_{\bm{q}s}} \cdot {{\vec J}^{(1)}}({{\vec \delta }_3})} \right)\left( {\vec q \cdot {{\vec \delta }_3}} \right) +{2\left( {{{\hat e}_{\bm{q}s}} \cdot {{\vec K}^{(1)}}({{\vec \delta }_1})} \right)\left( {\vec q \cdot {{\vec \delta }_1}} \right)}\Big]\\
& = i\frac{{4\pi n}}{3}\left( {{{\hat e}_{\bm{q}s}} \cdot {{\vec J}^{(1)}} + {{\hat e}_{\bm{q}s}} \cdot {{\vec K}^{(1)}}} \right),
\end{align*}
\begin{align*}
&B'({\bm{k}},{\bm{q}}) =i\left( {{{\hat e}_{\bm{q}s}} \cdot {{\vec J}^{(1)}}({{\vec \delta }_1})} \right)\left( {\vec q \cdot {{\vec \delta }_1}} \right){e^{ - i\vec k \cdot {{\vec \delta }_1}}}\\
&= i\frac{{2\pi n}}{3}\left( {{{\hat e}_{\bm{q}s}} \cdot {{\vec J}^{(1)}}} \right){\eta ^{ - 2}},
\end{align*}
\begin{align*}
&C'({\bm{k}},{\bm{q}}) = i\Big[{\left( {{{\hat e}_{\bm{q}s}} \cdot {{\vec K}^{(1)}}({{\vec \delta }_3})} \right)\left( {\vec q \cdot {{\vec \delta }_3}} \right){e^{ - i\vec k \cdot {{\vec \delta }_3}}}}  \\
&- \left( {{{\hat e}_{\bm{q}s}} \cdot {{\vec K}^{(1)}}({{\vec \delta }_2})} \right)\left( {\vec q \cdot {{\vec \delta }_2}} \right){e^{ - i\vec k \cdot {{\vec \delta }_2}}}
\Big]\\
&=- 2im\pi \left( {{{\hat e}_{\bm{q}s}} \cdot {{\vec K}^{(1)}}} \right)\zeta \eta,  
\end{align*}
\begin{align*}
&D'({\bm{k}},{\bm{q}}) =i\Bigg( \left[ {\left( {{{\hat e}_{\bm{q}s}} \cdot {{\vec J}^{(1)}}({{\vec \delta }_2})} \right) + \left( {{{\hat e}_{\bm{q}s}} \cdot {{\vec K}^{(1)}}({{\vec \delta }_2})} \right)} \right]\\
&\left( {\vec q \cdot {{\vec \delta }_2}} \right){e^{ - i\vec k \cdot {{\vec \delta }_2}}}+\left[ {\left( {{{\hat e}_{\bm{q}s}} \cdot {{\vec J}^{(1)}}({{\vec \delta }_3})} \right) + \left( {{{\hat e}_{\bm{q}s}} \cdot {{\vec K}^{(1)}}({{\vec \delta }_3})} \right)} \right] \\
&\left( {\vec q \cdot {{\vec \delta }_3}} \right) {e^{ - i\vec k \cdot {{\vec \delta }_3}}}\Bigg)=- i\frac{{2\pi n}}{3}\left( {{{\hat e}_{\bm{q}s}} \cdot {{\vec J}^{(1)}} + {{\hat e}_{\bm{q}s}} \cdot {{\vec K}^{(1)}}} \right)\zeta \eta,   
\end{align*} 
where 
\begin{equation*}
 {\bm{q}} = m\frac{{2\pi }}{a}{{{\bm{\hat e}}}_X} + n\frac{{2\pi }}{b}{{{\bm{\hat e}}}_Y},\;\; a=\alpha \sqrt{3},\; b=3\alpha,\;\;m,n \in \mathbb{Z},
 \end{equation*}
and also, $\zeta  = {e^{i\pi h}} = {\zeta ^{ - 1}}$ and $\eta  = {e^{ik\pi /3 }}$. Notice that in the calculation of the parameters  $B'(\bm{k},\bm{q})$, $C'(\bm{k},\bm{q})$ and $D'(\bm{k},\bm{q})$ above, the substitution $\bm{k} \to \bm{-k}$ implies the substitution $(h,k) \to (-h,-k)$ (i.e. switch the sign of the magnon wavector components; see the definitions prior to Eqs.\eqref{eq:zigzagdispersions1} and \eqref{eq:zigzagdispersions2}), and further, it was assumed that the exchange couplings $\vec J^{(1)}$ and $\vec K^{(1)}$ are bond independent, as a result of which the bond direction dependence was dropped. 

\subsection{Stripy phase}
For the stripy phase the magnetic unit cell is again defined by the gray rectangle of sides $a$ and $b$ shown in Fig. \ref{fig:zigzagcell}, consisting of four magnetic moments A, B, C and D, with A and B pointing along the positive X-axis, and C and D pointing along the negative X-axis. Choosing the positive spin quantization axis along the negative X-axis again, at the sites A and B we employ the bosonization given by the Eqs.\eqref{eq:negativeHPoperators1}-\eqref{eq:negativeHPoperators3}, while at the sites C and D we employ the bosonization given by the Eqs.\eqref{eq:positiveHPoperators1}-\eqref{eq:positiveHPoperators3}.
\begin{figure}
    \centering
    \includegraphics[scale=0.32]{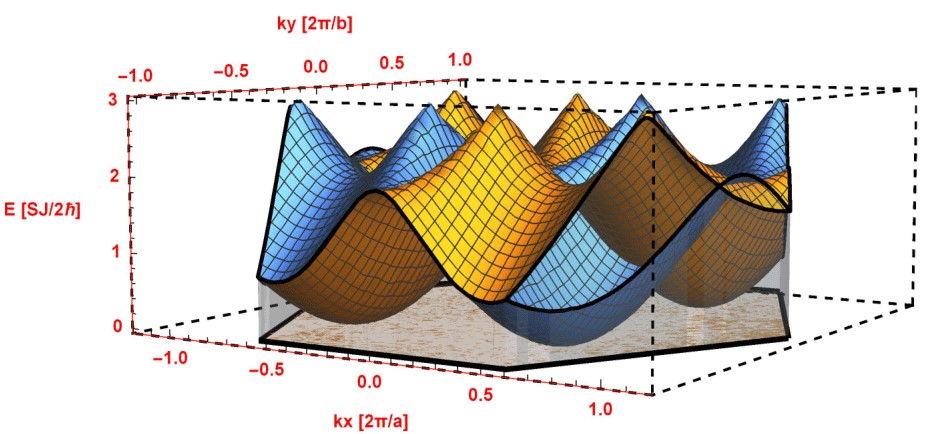}
    \caption{(Color online) Lower spin wave dispersion relations of the stripy phase, as given by Eqs.\eqref{eq:stripydispersions1}. The yellow surface corresponds to $\omega_1(\bm{k})$ and the blue surface to $\omega_2(\bm{k})$. Notice that the magnon wavevector components $k_x$ and $k_y$ are measured in units of $\frac{2\pi}{a}$ and $\frac{2\pi}{b}$ respectively and the spin wave energy is measured in units of $\frac{SJ}{2\hbar}$. The shaded hexagon within the Oxy plane is the 1BZ of the honeycomb lattice. The plot is for $K/J=-1$ and $\alpha = 2\pi/3$.}
    \label{fig:stripyspinwave1}
\end{figure}
The nn bonds that interact through the Hamiltonian of Eq.~\eqref{eq:HKham} as well as the Kitaev couplings are the same as in the case of the zig-zag phase. Using the representation of the $x$- and $y$-spin components in terms of the ladder spin operators to write the total Hamiltonian in terms of the $S_i^{||}$, $S_i^+$ and $S_i^-$ operators, performing the bosonization as elaborated above, and Fourier transforming according to the convention of Eqs.\eqref{Fourieroperators1} and \eqref{Fourieroperators2}, ones finds the classical ground state energy ${\cal H}_{classical}=\frac{{N{S^2}}}{2}\left( {-J + 2K} \right) $, and the following $\textit{spin wave mode matrix}$ $M(\bm{k})$ (reference to Eq.\eqref{eq:spinwave} and the notation thereof):
\begin{widetext}
\begin{equation}
{{M} ({\bm{k}}) =\left[ {\begin{array}{*{20}{c}}
  A&{{B^*(\bm{k})}}&0&{{C^*(\bm{k})}}&0&0&0&{{D^*(\bm{k})}} \\ 
  B(\bm{k})&A&C(\bm{k})&0&0&0&D(\bm{k})&0 \\ 
  0&{{C^*(\bm{k})}}&A&{{B^*(\bm{k})}}&0&{{D^*(\bm{k})}}&0&0 \\ 
  C(\bm{k})&0&B(\bm{k})&A&D(\bm{k})&0&0&0 \\ 
  0&0&0&{{D^*(\bm{k})}}&A&{{B^*(\bm{k})}}&0&{{C^*(\bm{k})}} \\ 
  0&0&D(\bm{k})&0&B(\bm{k})&A&C(\bm{k})&0 \\ 
  0&{{D^*(\bm{k})}}&0&0&0&{{C^*(\bm{k})}}&A&{{B^*(\bm{k})}} \\ 
  D(\bm{k})&0&0&0&C(\bm{k})&0&B(\bm{k})&A 
\end{array}} \right]},
\end{equation}
\end{widetext}
where we defined the following parameters
\begin{align*}
  &A = J - 2K,   \\
  &B({\bm{k}}) = J{e^{ - i\vec k \cdot {{\vec \delta }_1}}} = J{\eta ^{ - 2},}   \\
  &C({\bm{k}}) = 
 K\left({e^{ - i\vec k \cdot {{\vec \delta }_3}}} - {e^{ - i\vec k \cdot {{\vec \delta }_2}}}\right) = 2iK\eta \sin (\pi h),   \\
  &D({\bm{k}}) = \left( {J + K} \right) \left( {e^{ - i\vec k \cdot {{\vec \delta }_3}}} 
  + {e^{ - i\vec k \cdot {{\vec \delta }_2}}} \right)
 = 2(J + K)\eta \cos(\pi h). 
\end{align*} 
and as previously it is 
\begin{equation}
\zeta  = {e^{i\pi h}} = {\zeta ^{ - 1}}\;\;\;({\zeta ^2} = 1 = \zeta {\zeta ^{ - 1}}),\;\;\;\eta  = {e^{ik\pi /3 }}. \nonumber
\end{equation}

\begin{figure}
    \centering
    \includegraphics[scale=0.3]{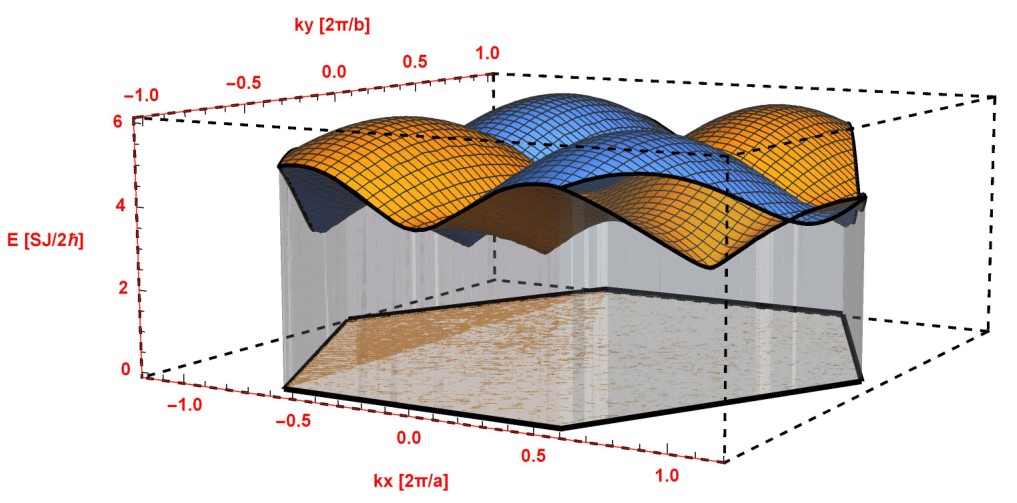}
    \caption{(Color online) Upper spin wave dispersion relations of the stripy phase, as given by Eqs.\eqref{eq:stripydispersions2}. The yellow surface corresponds to $\omega_3(\bm{k})$ and the blue surface to $\omega_4(\bm{k})$. Notice that the magnon wavevector components $k_x$ and $k_y$ are measured in units of $\frac{2\pi}{a}$ and $\frac{2\pi}{b}$ respectively and the spin wave energy is measured in units of $\frac{SJ}{2\hbar}$. The shaded hexagon within the Oxy plane is the 1BZ of the honeycomb lattice. The plot is for $K/J=-1$ and $\alpha = 2\pi/3$.}
    \label{fig:stripyspinwave2}
\end{figure}

Diagonalizing the dynamical matrix $D=I_-M$ as described in Eq.\eqref{eq:dynamicalmatrix} in Sec.\ref{sec:spin_waves}, we obtain the following $\textit{magnon normal modes}$:

\begin{eqnarray}
&\label{eq:stripydispersions1}
  \omega _1  = \sqrt {{\Omega _1} - \sqrt {{\Omega _2}} },\;\;\;
  \omega _2  = \sqrt {{\Omega _3} - \sqrt {{\Omega _4}} }\\
&\label{eq:stripydispersions2}
  \omega _3  = \sqrt {{\Omega _1} + \sqrt {{\Omega _2}} },\;\;\;
  \omega _4  = \sqrt {{\Omega _3} + \sqrt {{\Omega _4}} },
\end{eqnarray} 
where the following parameters were used
\begin{align*}
  &{\Omega _1} = {A^2} - {\left| D \right|^2} + {\left| {B - C} \right|^2},\\
 & {\Omega _2} = 4{\left|A (B-C) \right|^2} - {\left| {D({B^*} - {C^*}) - {D^*}(B - C)} \right|^2},   \\
 & {\Omega _3} = {A^2} - {\left| D \right|^2} + {\left| {B + C} \right|^2},\\
 &{\Omega _4} = 4{\left|A (B+C) \right|^2}  - {\left| {D({B^*} + {C^*}) - {D^*}(B + C)} \right|^2}. 
\end{align*}
The spin wave dispersions of Eq.\eqref{eq:stripydispersions1} are plotted in Fig.\ref{fig:stripyspinwave1} and those of Eq.\eqref{eq:stripydispersions2} are plotted in Fig.\ref{fig:stripyspinwave2}. As can be seen from Figs.\ref{fig:stripyspinwave1} and \ref{fig:stripyspinwave2} (vertical axis), the lower and the upper magnon bands are well-separated in energy from each other. At low enough temperatures, as far as the magnon-phonon interaction is concerned, only the parts of the spin wave spectra around the $\textit{spin wave valleys}$, and in this case the lower magnon bands are of interest. The lower energy magnon valley  $\bm{k}$-space positions are found from the conditions that  $\omega_i=0$, $i={1,2}$, which can be solved as was detailed in the previous section. One can check that the spin wave dispersions of  Eqs.\eqref{eq:stripydispersions1} and \eqref{eq:stripydispersions2}  have the symmetry properties: ${\omega_i}(k_x,k_y) = {\omega_i}(-k_x,-k_y)\;\;(\textit{time reversal symmetry})$, as well as ${\omega_i}(k_x,-k_y) = {\omega_i}(-k_x,k_y)$, for $i = 1,2,3,4 $, which can further be employed to simplify the calculations.
As far as the magnon phonon scattering matrix is concerned, the $\Lambda'(\bm{k},\bm{q})$ matrix on the RHS of Eq.\eqref{eq:scatter} has the following form 
\begin{widetext}
\begin{equation}
{\Lambda' } ({\bm{k}},{\bm{q}}) =\left[ {\begin{array}{*{20}{c}}
  {A'(\bm{q})}&{B'(\bm{-k},\bm{q})}&0&{C'(\bm{-k},\bm{q})}&0&0&0&{D'(\bm{-k},\bm{q})} \\ 
  {B'(\bm{k},\bm{q})}&{A'(\bm{q})}&{C'(\bm{k},\bm{q})}&0&0&0&{D'(\bm{k},\bm{q})}&0 \\ 
  0&{C'(\bm{-k},\bm{q})}&{A'(\bm{q})}&{B'(\bm{-k},\bm{q})}&0&{D'(\bm{-k},\bm{q})}&0&0 \\ 
  {C'(\bm{k},\bm{q})}&0&{B'(\bm{k},\bm{q})}&{A'(\bm{q})}&{D'(\bm{k},\bm{q})}&0&0&0 \\ 
  0&0&0&{D'(\bm{-k},\bm{q})}&{A'(\bm{q})}&{B'(\bm{-k},\bm{q})}&0&{C'(\bm{-k},\bm{q})} \\ 
  0&0&{D'(\bm{k},\bm{q})}&0&{B'(\bm{k},\bm{q})}&{A'(\bm{q})}&{C'(\bm{k},\bm{q})}&0 \\ 
  0&{D'(\bm{-k},\bm{q})}&0&0&0&{C'(\bm{-k},\bm{q})}&{A'(\bm{q})}&{B'(\bm{-k},\bm{q})} \\ 
  {D'(\bm{k},\bm{q})}&0&0&0&{C'(\bm{k},\bm{q})}&0&{B'(\bm{k},\bm{q})}&{A'(\bm{q})}  
\end{array}} \right],
\end{equation}
\end{widetext}
where the parameters $B'(\bm{k},\bm{q})$, $C'(\bm{k},\bm{q})$ and $D'(\bm{k},\bm{q})$ are defined exactly as in the zig-zag phase, with the following modification for the $A'(\bm{q})$ parameter
\begin{equation*}
A'({\bm{q}})  
 = -i\frac{{4\pi n}}{3}\left( {{{\hat e}_{\bm{q}s}} \cdot {{\vec J}^{(1)}} + {{\hat e}_{\bm{q}s}} \cdot {{\vec K}^{(1)}}} \right),
\end{equation*}
and further,
\begin{equation*}
 {\bm{q}} = m\frac{{2\pi }}{a}{{{\bm{\hat e}}}_X} + n\frac{{2\pi }}{b}{{{\bm{\hat e}}}_Y},\;\; a=\alpha \sqrt{3},\; b=3\alpha,\;\;m,n \in \mathbb{Z},
 \end{equation*}
 $\zeta  = {e^{i\pi h}} = {\zeta ^{ - 1}}$ and $\eta  = {e^{ik\pi /3 }}$. Notice again that in the calculation of the parameters $B'(\bm{k},\bm{q})$, $C'(\bm{k},\bm{q})$ and $D'(\bm{k},\bm{q})$ above, the substitution $\bm{k} \to \bm{-k}$ implies the substitution $(h,k) \to (-h,-k)$, and it was assumed that the exchange couplings $J$ and $K$ as well as the couplings $\vec J^{(1)}$ and $\vec K^{(1)}$ are bond independent, as a result of which the bond direction dependence was dropped. 


\subsection{N\'eel phase}

For the N\'eel phase the magnetic unit cell coincides with the chemical unit cell defined by the parallelogram of sides $\bm{t_1}$ and $\bm{t_2}$ (see Fig. \ref{fig:zigzagcell}), and consists of two magnetic moments A, B, with A pointing along the positive X-axis, and B pointing along the negative X-axis. Choosing the positive spin quantization axis along the negative X-axis again, at the site A we employ the bosonization given by the Eqs.\eqref{eq:negativeHPoperators1}-\eqref{eq:negativeHPoperators3}, while at the site B we employ the bosonization given by the Eqs.\eqref{eq:positiveHPoperators1}-\eqref{eq:positiveHPoperators3}.
\begin{figure}
    \centering
    \includegraphics[scale=0.32]{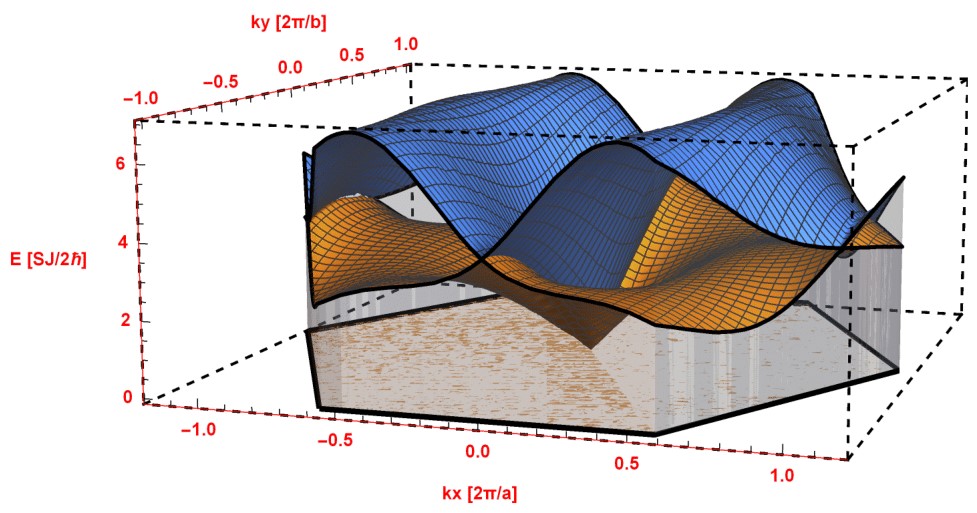}
    \caption{(Color online) Spin wave dispersion relations of the N\'eel phase, as given by Eqs.\eqref{eq:Neelspectra1}. The yellow surface corresponds to $\omega_1(\bm{k})$ and the blue surface to $\omega_2(\bm{k})$. Notice that the magnon wavevector components $k_x$ and $k_y$ are measured in units of $\frac{2\pi}{a}$ and $\frac{2\pi}{b}$ respectively and the spin wave energy is measured in units of $\frac{SJ}{2\hbar}$. The shaded hexagon within the Oxy plane is the 1BZ of the honeycomb lattice. The plot is for $K/J=1$ and $\alpha = 2\pi/3$.}
    \label{fig:neelspinwave}
 \end{figure}
 The bond dependent Kitaev couplings are defined as in the zig-zag phase, except that now only the nn bonds at sites A and B are taken into account leading to a total of three bonds. Using the representation of the $x$- and $y$-spin components in terms of the ladder spin operators to write the total Hamiltonian in terms of the $S_i^{||}$, $S_i^+$ and $S_i^-$ operators, performing the bosonization as elaborated above, and Fourier transforming according to the convention of Eqs.\eqref{Fourieroperators1} and \eqref{Fourieroperators2}, ones finds the classical ground state energy ${\cal H}_{classical}=-\frac{{N{S^2}}}{2}\left( {3J + 2K} \right) $, and the following $\textit{spin wave mode matrix}$ $M(\bm{k})$ (reference to Eq.\eqref{eq:spinwave} and the notation thereof): 
\begin{equation}
{M} ({\bm{k}}) =\left[
  {\begin{array}{*{20}{c}}
A&C(\bm{k})&0&B(\bm{k})\\
C^*(\bm{k})&A&B^*(\bm{k})&0\\
0&B(\bm{k})&A&C(\bm{k})\\
B^*(\bm{k})&0&C^*(\bm{k})&A
\end{array}} \right],
\end{equation}
where we defined the following parameters
 \begin{align*}
  &A = 3J + 2K,   \\
  &B({\bm{k}}) = J({{\vec \delta }_1}){e^{ - i\vec k \cdot {{\vec \delta }_1}}} + \left( {J({{\vec \delta }_3}) + K({{\vec \delta }_3})} \right){e^{ - i\vec k \cdot {{\vec \delta }_3}}} \\
  &+ \left( {J({{\vec \delta }_2}) 
 + K({{\vec \delta }_2})} \right){e^{ - i\vec k \cdot {{\vec \delta }_2}}}  = J{\eta ^{ - 2}+2(J + K)\eta \cos(\pi h),}   \\
  &C({\bm{k}}) = K({{\vec \delta }_3}){e^{ - i\vec k \cdot {{\vec \delta }_3}}}-K({{\vec \delta }_2}){e^{ - i\vec k \cdot {{\vec \delta }_2}}} = 2iK\eta \sin (\pi h),   
\end{align*} 
in conjunction with the definitions  
\begin{equation}
\zeta  = {e^{i\pi h}} = {\zeta ^{ - 1}}\;\;\;({\zeta ^2} = 1 = \zeta {\zeta ^{ - 1}}),\;\;\;\eta  = {e^{ik\pi /3 }}. \nonumber
\end{equation}

Diagonalizing the dynamical matrix $D=I_-M$ as described in Eq.\eqref{eq:dynamicalmatrix} in Sec.\ref{sec:spin_waves}, we obtain the following $\textit{magnon normal modes}$:
\begin{eqnarray}
\label{eq:Neelspectra1}
  \omega _1  = \sqrt {{\Omega _1} - \sqrt {{\Omega _2}} }, \;
  \omega _2  = \sqrt {{\Omega _1} + \sqrt {{\Omega _2}} },
\end{eqnarray} 
where the following parameters were used
\begin{align*}
 &{\Omega _1} = {A^2} - {\left| B \right|^2} + {\left| { C} \right|^2},\\ 
 & {\Omega _2} = 4{A^2 \left| C \right|^2} + (B^*C - C^*B)^2.   
 \end{align*}
 The spin wave dispersions of Eq.\eqref{eq:Neelspectra1} are plotted in Fig.\ref{fig:neelspinwave}. At low enough temperatures, as far as the magnon-phonon interaction is concerned, only the parts of the spin wave spectra around the $\textit{spin wave valleys}$ are of interest, which in this case are located at the $\Gamma$-point of the 1BZ (as opposed to the previous phases). The exact magnon valley  $\bm{k}$-space positions are found from the conditions that  $\omega_i=0$, $i={1,2}$, which can be solved as was detailed in the previous sections. One can check that the spin wave dispersions of Eq.\eqref{eq:Neelspectra1}  have the symmetry properties: ${\omega_i}(k_x,k_y) = {\omega_i}(-k_x,-k_y)\;\;(\textit{time reversal symmetry})$ as well as  ${\omega_i}(k_x,-k_y) = {\omega_i}(-k_x,k_y)$, for $i = 1,2 $, which can further be employed to simplify the calculations.
As far as the magnon phonon scattering matrix is concerned, the $\Lambda'(\bm{k},\bm{q})$ matrix on the RHS of Eq.\eqref{eq:scatter} has the following form 
\begin{equation}
{\Lambda' } ({\bm{k}},{\bm{q}}) =\left[ {\begin{array}{*{20}{c}}
  {A'(\bm{q})}& {C'(\bm{k},\bm{q})}&0&{B'(\bm{k},\bm{q})}\\ 
   {C'(-\bm{k},\bm{q})}&{A'(\bm{q})}&{B'(\bm{-k},\bm{q})}&0 \\ 
  0& {B'(\bm{k},\bm{q})}&{A'(\bm{q})}& {C'(\bm{k},\bm{q})} \\ 
  {B'(-\bm{k},\bm{q})}&0&{C'(-\bm{k},\bm{q})}&{A'(\bm{q})}
\end{array}} \right],
\end{equation}
where the parameters $A'(\bm{q})$, $B'(\bm{k},\bm{q})$ and $C'(\bm{k},\bm{q})$ are defined as below
\begin{align*}
&A'(\bm q)= i \frac{4 \pi}{3} (\hat{e}_{\bm{q}s} \cdot \vec{K}^{(1)}),\\
&B'(\bm{k}, \bm {q})= i\frac{{2\pi n}}{3}\left( {{{\hat e}_{\bm{q}s}} \cdot {{\vec J}^{(1)}}} \right){\eta ^{ - 2}}\\
&- i\frac{{2\pi n}}{3}\left( {{{\hat e}_{\bm{q}s}} \cdot {{\vec J}^{(1)}} + {{\hat e}_{\bm{q}s}} \cdot {{\vec K}^{(1)}}} \right)\zeta \eta,\\
&C'(\bm{k}, \bm {q})=- 2im\pi \left( {{{\hat e}_{\bm{q}s}} \cdot {{\vec K}^{(1)}}} \right)\zeta \eta,
\end{align*}
and further,
\begin{equation*}
 {\bm{q}} = m\frac{{2\pi }}{a}{{{\bm{\hat e}}}_X} + n\frac{{2\pi }}{b}{{{\bm{\hat e}}}_Y},\;\; a=\alpha \sqrt{3},\; b=3\alpha,\;\;m,n \in \mathbb{Z},
 \end{equation*}
 $\zeta  = {e^{i\pi h}} = {\zeta ^{ - 1}}$ and $\eta  = {e^{ik\pi /3 }}$. Notice that in the calculation of the parameters $B'(\bm{k},\bm{q})$ and $C'(\bm{k},\bm{q})$ above, the substitution $\bm{k} \to \bm{-k}$ implies the substitution $(h,k) \to (-h,-k)$, and it was assumed that the exchange couplings $J$ and $K$ as well as the couplings $\vec J^{(1)}$ and $\vec K^{(1)}$ are bond independent, as a result of which the bond direction dependence was dropped. 

\subsection{Ferromagnetic phase}
For the ferromagnetic phase the magnetic unit cell again coincides with the chemical unit cell defined by the parallelogram of sides $\bm{t_1}$ and $\bm{t_2}$ (see Fig. \ref{fig:zigzagcell}), and consists of two magnetic moments A, B, with A and B both pointing along the positive X-axis. Choosing the positive spin quantization axis along the negative X-axis again, at both sites we employ the bosonization given by the Eqs.\eqref{eq:negativeHPoperators1}-\eqref{eq:negativeHPoperators3}.
The nn bonds that interact through the Hamiltonian of Eq.~\eqref{eq:HKham} as well as the definition of the Kitaev couplings are the same as in the case of the N\'eel phase. Using the representation of the $x$- and $y$-spin components in terms of the ladder spin operators to write the total Hamiltonian in terms of the $S_i^{||}$, $S_i^+$ and $S_i^-$ operators, performing the bosonization as elaborated above, and Fourier transforming according to the convention of Eqs.\eqref{Fourieroperators1} and \eqref{Fourieroperators2}, ones finds the classical ground state energy ${\cal H}_{classical}=\frac{{N{S^2}}}{2}\left( {3J + 2K} \right) $, and the following $\textit{spin wave mode matrix}$ $M(\bm{k})$ (reference to Eq.\eqref{eq:spinwave} and the notation thereof): 
\begin{equation}
{M} ({\bm{k}}) =\left[
  {\begin{array}{*{20}{c}}
A&B(\bm{k})&0&C(\bm{k})\\
B^*(\bm{k})&A&C^*(\bm{k})&0\\
0&C(\bm{k})&A&B(\bm{k})\\
C^*(\bm{k})&0&B^*(\bm{k})&A
\end{array}} \right],
\end{equation}
where the parameters $B(\bm{k})$ and $C(\bm{k})$ are defined exactly as in the N\'eel phase, except for the parameter $A$ which is modified as below
\begin{equation}
\label{eq:alphaFM}
A= - 3J - 2K.
\end{equation}
\begin{figure}
    \centering
    \includegraphics[scale=0.36]{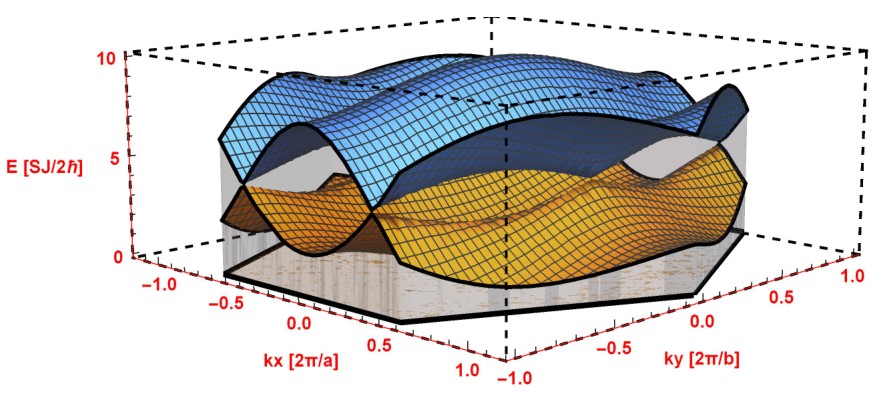}
    \caption{(Color online) Spin wave dispersion relations of the ferromagnetic phase, as given by Eqs.\eqref{eq:fmspectra1}. The yellow surface corresponds to $\omega_1(\bm{k})$ and the blue surface to $\omega_2(\bm{k})$. Notice that the magnon wavevector components $k_x$ and $k_y$ are measured in units of $\frac{2\pi}{a}$ and $\frac{2\pi}{b}$ respectively and the spin wave energy is measured in units of $\frac{SJ}{2\hbar}$. The shaded hexagon within the Oxy plane is the 1BZ of the honeycomb lattice. The plot is for $K/J=1$ and $\alpha = 2\pi/3$.}
    \label{fig:fmspinwave}
\end{figure}

Diagonalizing the dynamical matrix $D=I_-M$ as described in Eq.\eqref{eq:dynamicalmatrix} in Sec.\ref{sec:spin_waves}, we obtain the following $\textit{magnon normal modes}$:
\begin{eqnarray}
\label{eq:fmspectra1}
  \omega _1  = \sqrt {{\Omega _1} - \sqrt {{\Omega _2}} }, \;
  \omega _2  = \sqrt {{\Omega _1} + \sqrt {{\Omega _2}} },
\end{eqnarray} 
where the following parameters were used
\begin{align*}
 &{\Omega _1} = {A^2} - {\left| C \right|^2} + {\left| { B} \right|^2},\\ 
 & {\Omega _2} = 4{A^2 \left| B \right|^2} + (B^*C - C^*B)^2.   
 \end{align*} 
The spin wave dispersions of Eq.\eqref{eq:fmspectra1} are plotted in Fig.\ref{fig:fmspinwave}. At low enough temperatures, as far as the magnon-phonon interaction processes are concerned, only the part of the lower spin wave spectum around the $\textit{spin wave valley}$ is of interest, which in this case is again located at the $\Gamma$-point of the 1BZ. The exact magnon valley $\bm{k}$-space position is found from the condition that $\omega_i=0$, $i={1}$, which can be solved as was detailed in the previous sections. One can check that the spin wave dispersions of the ferromagnetic phase as well have the symmetry properties: ${\omega_i}(k_x,k_y) = {\omega_i}(-k_x,-k_y)\;\;(\textit{time reversal symmetry})$ as well as ${\omega_i}(k_x,-k_y) = {\omega_i}(-k_x,k_y)$, for $i = 1,2 $. It should be stressed that the spin wave dispersions of both the N\'eel and the ferromagnetic phase, in Figs.\ref{fig:neelspinwave} and \ref{fig:fmspinwave} respectively, are not exactly isotropic around the respective spin wave valleys. 
Lastly, for the magnon phonon scattering matrix $\Lambda'(\bm{k},\bm{q})$ of the RHS of Eq.\eqref{eq:scatter} it is 
\begin{equation}
{\Lambda' } ({\bm{k}},{\bm{q}}) =\left[ {\begin{array}{*{20}{c}}
  {A'(\bm{q})}& {B'(\bm{k},\bm{q})}&0&{C'(\bm{k},\bm{q})}\\ 
   {B'(-\bm{k},\bm{q})}&{A'(\bm{q})}&{C'(\bm{-k},\bm{q})}&0 \\ 
  0& {C'(\bm{k},\bm{q})}&{A'(\bm{q})}& {B'(\bm{k},\bm{q})} \\ 
  {C'(-\bm{k},\bm{q})}&0&{B'(-\bm{k},\bm{q})}&{A'(\bm{q})}
\end{array}} \right],
\end{equation}
where the parameters $B'(\bm{k},\bm{q})$ and $C'(\bm{k},\bm{q})$ are defined exactly as in the N\'eel phase, with the following modification for the $A'(\bm{q})$ parameter
\begin{align*}
&A'(\bm q)= -i \frac{4 \pi}{3} (\hat{e}_{\bm{q}s} \cdot \vec{K}^{(1)}),
\end{align*}
and further,
\begin{equation*}
 {\bm{q}} = m\frac{{2\pi }}{a}{{{\bm{\hat e}}}_X} + n\frac{{2\pi }}{b}{{{\bm{\hat e}}}_Y},\;\; a=\alpha \sqrt{3},\; b=3\alpha,\;\;m,n \in \mathbb{Z},
 \end{equation*}
 $\zeta  = {e^{i\pi h}} = {\zeta ^{ - 1}}$ and $\eta  = {e^{ik\pi /3 }}$. It is again noticed that in the calculation of the parameters  $B'(\bm{k},\bm{q})$ and $C'(\bm{k},\bm{q})$ above, the substitution $\bm{k} \to \bm{-k}$ implies the substitution $(h,k) \to (-h,-k)$, and it was assumed that the exchange couplings $J$ and $K$ as well as the couplings $\vec J^{(1)}$ and $\vec K^{(1)}$ are bond independent, as a result of which the bond direction dependence was dropped. 
\\
\\
\\

\onecolumngrid

\section{\label{sec:level5} Boltzmann kinetic equation and derivation of thermal conductivity formula in the relaxation time approximation}

Consider the phase space of a multi-particle system of non-interacting particles, more generally non-interacting in the $\textit{mean field}$ sense. For such a case, instead of the multi-particle distribution function one can recourse to the so-called $\textit{reduced distribution functions}$ \cite{Pottier:oxf2010}, and more specifically to the $\textit{single-particle distribution function}$ $f({\bm{r}},{\bm{q}},t)$ without introducing any further approximations. Let us now focus on the motion of the particles whose phase space coordinates lie  within the volume $d{\bm{r}}d{\bm{q}}$ around the phase space point  
$({\bm{r}},{\bm{q}})$ at time $t$. If no collisions occur, then at time $t+dt$ the phase space coordinates of all those particles flow into the region $d{\bm{r}}'d{\bm{q}}'$  around the point
$({\bm{r}}',{\bm{q}}')$, where obviously
${\bm{r}}' = {\bm{r}} + {\bm{\dot r}}dt$  and
${\bm{q}}' = {\bm{q}} + {\bm{\dot q}}dt$. Conservation of the number of particles (since no collisions occur) dictates that 
\begin{equation}
f({\bm{r}},{\bm{q}},t)d{\bm{r}}d{\bm{q}} = f({\bm{r}}',{\bm{q}}',t + dt)d{\bm{r}}'d{\bm{q}}' = f({\bm{r}} + {\bm{\dot r}}dt,{\bm{q}} + {\bm{\dot q}}dt,t + dt)d{\bm{r}}'d{\bm{q}}' ,\nonumber
\end{equation}
where $f({\bm{r}},{\bm{q}},t)$  is the single-particle dynamical phase-space distribution function. Liouville's theorem states that 
$d{\bm{r}}d{\bm{q}} = d{\bm{r}}'d{\bm{q}}'$, implying that
\begin{equation}
\label{eq:Liouville}
f({\bm{r}} + {\bm{\dot r}}dt,{\bm{q}} + {\bm{\dot q}}dt,t + dt) = f({\bm{r}},{\bm{q}},t).
\end{equation}
Furthemore,
\begin{equation}
\label{eq:df}
f({\bm{r}} + {\bm{\dot r}}dt,{\bm{q}} + {\bm{\dot q}}dt,t + dt) = f({\bm{r}},{\bm{q}},t) + \frac{{\partial f}}{{\partial {\bm{r}}}}{\bm{\dot r}}dt + \frac{{\partial f}}{{\partial {\bm{q}}}}{\bm{\dot q}}dt + \frac{{\partial f}}{{\partial t}}dt.
\end{equation}
Combining Eqs.\eqref{eq:Liouville}, \eqref{eq:df} we get
\begin{equation}
\label{eq:totaldistributionfunc}
\frac{{df}}{{dt}} = \frac{{f({\bm{r}} + {\bm{\dot r}}dt,{\bm{q}} + {\bm{\dot q}}dt,t + dt) - f({\bm{r}},{\bm{q}},t)}}{{dt}} = {\bm{\dot r}}\frac{{\partial f}}{{\partial {\bm{r}}}} + {\bm{\dot q}}\frac{{\partial f}}{{\partial {\bm{q}}}} + \frac{{\partial f}}{{\partial t}} = 0. 
\end{equation}
Now, if collisions do occur during the infinitesimal time interval $dt$, some particles are scattered out (of the aforementioned multi-particle distribution function) whereas other particles are scattered in (the aforementioned multi-particle distribution function), upon flowing from the phase space point 
$({\bm{r}},{\bm{q}})$ to the phase space point 
$({\bm{r}}',{\bm{q}}')$,  infinitesimally far away (within the phase space). As a result of it, the single-particle dynamical phase space distribution function does not satisfy Eq.\eqref{eq:totaldistributionfunc}, but instead it is (reducing the inscattering and outscattering from the multi-particle distribution function to a probability of inscattering and outscattering from the single-particle distribution function)
\begin{equation}
\label{eq:collisionterm}
\frac{{f({\bm{r}} + {\bm{\dot r}}dt,{\bm{q}} + {\bm{\dot q}}dt,t + dt) - f({\bm{r}},{\bm{q}},t)}}{{dt}} =\left( \frac{{df}}{{dt}} \right)_{in}-\left( \frac{{df}}{{dt}} \right)_{out} \equiv \left( \frac{{df}}{{dt}} \right)_{coll}, 
\end{equation}
where the rightmost term accounts for the total change in the single-particle distribution function due to inscattering and outscattering processes, and is the so-called $\textit{collision term}$. Combining Eqs. \eqref{eq:totaldistributionfunc} and \eqref{eq:collisionterm}, to linear order in  $dt$ we get 
\begin{align}
  {\bm{\dot r}}  \frac{{\partial f}}{{\partial {\bm{r}}}}+ {\bm{\dot q}}\frac{{\partial f}}{{\partial {\bm{q}}}} + \frac{{\partial f}}{{\partial t}}= \left( \frac{{df}}{{dt}} \right)_{coll}, 
\end{align} 
and this is the so-called $\textit{Boltzmann kinetic equation}$. Now, let us apply the Boltzmann kinetic equation to the problem of heat transport. Let us focus on the low energy lattice degrees of freedom, which in the language of second quanti-zation can be treated as non-interacting quasiparticles called $\textit{phonons}$, and derive an expression for the thermal conductivity tensor. 

When a temperature gradient (slowly varying in space, and time \cite{Pottier:oxf2010} in general) is present, phonons can be treated within the semiclassical approximation, i.e. they can be described by a semiclassical distribution function (from now on called $\textit{phonon distribution function}$) whose dynamics obeys the Boltzmann kinetic equation. For heat transport, the phononic distribution function is actually non-uniform in real space only, due to the presence of a nonzero tempera-ture gradient. As a result, the equation that governs the phase space variations of the phonon distribution function ${f_s}({\bm{r}},{\bm{q}},t)$, for phonons of polarization  $s$, has the following form
\begin{equation}
{{\bm{v}}_s}({\bm{q}})\frac{{\partial {f_s}({\bm{r}},{\bm{q}},t)}}{{\partial {\bm{r}}}} +\frac{{\partial {f_s}({\bm{r}},{\bm{q}},t)}}{{\partial t}}={\left( \frac{df_s}{dt} \right)_{coll}}, 
\end{equation}
where ${{\bm{v}}_s}({\bm{q}})$ is the group velocity of phonons of polarization $s$, given by ${{\bm{v}}_s}({\bm{q}}) = {\nabla _{\bm{q}}}{\omega _s}({\bm{q}})$. Taking into account the fact that the spatial non-uniformity of the phonon distribution function comes through the spatial variation of the temperature, in the so-called $\textit{stationary or steady state}$ case, one finds that 
\begin{equation}
\label{eq:botzmann}
 {{\bm{v}}_s}({\bm{q}}) \cdot \nabla_{\bm{r}} T\frac{{\partial {f_s}({\bm{q}})}}{{\partial T}}={\left( \frac{df_s}{dt} \right)_{coll}}. 
\end{equation}
Eq.\eqref{eq:botzmann} is the $\textit{stationary Boltzmann equation}$ for phonons of polarization $s$.
As was mentioned previously, in the phonon-dominated regime heat is mostly carried by the phonons, which at low enough temperatures can be treated as non-interacting quasiparticles, which weakly interact with a bath which in this case is the magnons. Under those conditions we attempt to solve the stationary Boltzmann equation \eqref{eq:botzmann} within the so-called $\textit{relaxation time approximation}$, and the current situation can be treated similarly to the impurity scattering of the electrons. 

Quite generally, the collision term can be put into the following form ($A$ denotes the area, and our analysis is adjusted to 2D)
\begin{equation}
\label{eq:collterm}
\left( \frac{df_s}{dt} \right)_{coll}\equiv I[f_s]=\frac{1}{A}\sum_{\bm{q'}}\left(W_{\bm{q'}\to\bm{q}} - W_{\bm{q}\to\bm{q'}}  \right), 
\end{equation}
where $W_{\bm{q'}\to \bm{q}}$ denotes the probability per unit time for a phonon to be scattered from $\bm{q'}$ to $\bm{q}$ in a given scattering process, which encompasses not only microscopic probabilities for quantum transitions but also the single-particle distribution function itself. More specifically, if the quantum transition probability per unit time, denoted as $w_{\bm{q'}\to\bm{q}}$ for a phononic scattering process from the state of wavevector $\bm{q'}$ to a state of wavevector $\bm{q}$ is known (this last quantity is directly related to the $\textit{magnon-phonon scattering matrix}$ of the analysis of the main text), then the probability $W_{\bm{q'}\to\bm{q}}$ can be expressed directly in terms of the microscopic probability $w_{\bm{q'}\to\bm{q}}$ and $f_{s}(\bm{r},\bm{q},t)$ (actually $f_{s}(\bm{q})$ for the stationary case that is of interest here). Furthermore, from Eq.\eqref{eq:collterm} it is 
\begin{align}
\label{eq:colltermrelax}
& I[f_s]=\frac{1}{A}\sum_{\bm{q'}}\left(w_{\bm{q'}\to\bm{q}} \left(f_s(\bm{q})+1 \right) - w_{\bm{q}\to\bm{q'}} f_s(\bm{q})  \right)=\frac{1}{A}\sum_{\bm{q'}}w_{\bm{q'}\to\bm{q}} \left(f_s(\bm{q})+1 \right)-\frac{1}{A}\sum_{\bm{q'}} w_{\bm{q}\to\bm{q'}} f_s(\bm{q})  \nonumber \\
& \equiv I_{in}[f_s]-I_{out}[f_s],
\end{align}
where in the rightmost term of Eq.\eqref{eq:colltermrelax} the collision term is decomposed into two parts, one related to inscattering and the other related to outscattering processes. Under thermal equilibrium conditions the inscattering and the outscattering processes should compensate each other leading to the result 
\begin{equation}
\label{eq:equicond}
w_{\bm{q'}\to\bm{q}} \left(f_s^0(\bm{q})+1 \right) = w_{\bm{q}\to\bm{q'}} f_s^0(\bm{q}),
\end{equation}
where $f_s^0(\bm{q})$ is the equilibrium distribution function. Assuming that the applied temperature gradient is such that the departure of the single-particle distribution function from its equilibrium value is small, i.e. $f_s(\bm{q}) \approx f_s^0(\bm{q})+f_s^1(\bm{q})$, from Eqs.\eqref{eq:colltermrelax} and \eqref{eq:equicond} to lowest order it is   
\begin{align}
\label{eq:colltermrel}
& I[f_s]\approx -\Bigg(\frac{1}{A} \sum_{\bm{q'}}\Big(w_{\bm{q}\to\bm{q'}} - w_{\bm{q'}\to\bm{q}} \Big) \Bigg) f_s^1(\bm{q})=-\Bigg[ \frac{1}{A} \sum_{\bm{q'}}\Big(w_{\bm{q}\to\bm{q'}} - w_{\bm{q'}\to\bm{q}} \Big) \Bigg] \Big(f_s(\bm{q})-f_s^0(\bm{q})\Big),
\end{align}
where we define the so-called $\textit{relaxation time}$ as below
\begin{equation}
\label{eq:relaxtime}
\frac{1}{\tau_s(\bm{q})}= \frac{1}{A} \sum_{\bm{q'}}\left( w_{\bm{q}\to\bm{q'}} - w_{\bm{q'}\to\bm{q}}  \right).
\end{equation}
Notice that the result of Eq.\eqref{eq:relaxtime} per unit area is directly related to Eqs.\eqref{eq:phoreltimes} and \eqref{eq:phototalreltimes} that were derived in the phonon-dominated thermal transport regime. Notice also that the microscopic transition probabilities $w_{\bm{q}\to\bm{q'}}$ and $w_{\bm{q'}\to\bm{q}}$ do not necessarily balance each other (as happens in the problem of the elastic scattering of an electron from impurities), and more specifically, to ensure the $\textit{non-negativity}$ of the phonon relaxation time defined above it should be true that $w_{\bm{q}\to\bm{q'}} \geq  w_{\bm{q'}\to\bm{q}}$, and of course the quantity $\sum_{\bm{q'}}\Big( w_{\bm{q}\to\bm{q'}} - w_{\bm{q'}\to\bm{q}} \Big)$ should be bounded (not infinite). Under the aforementioned conditions, the weak interaction of phonons with the magnon bath (under a weak temperature gradient) can be described via the concept of the $\textit{phonon relaxation time}$. 

Combining Eqs.\eqref{eq:botzmann}, \eqref{eq:collterm}, \eqref{eq:colltermrel} and \eqref{eq:relaxtime} we get (using again the approximation of $f_s(\bm{q}) \approx f_s^0(\bm{q})+f_s^1(\bm{q})$)
\begin{equation}
\label{eq:botzmann2}
 {{\bm{v}}_s}({\bm{q}}) \cdot \nabla T\frac{{\partial {f_s^0}(\bm{q})}}{{\partial T}}+{{\bm{v}}_s}({\bm{q}}) \cdot \nabla T\frac{{\partial {f_s^1}(\bm{q})}}{{\partial T}}= - \frac{f_s^1{(\bm{q}})}{\tau_s(\bm{q})} \nonumber,
\end{equation}
and neglecting on the LHS (left hand side) the term that depends on $f_s^1(\bm{q})$ (as being smaller compared to the other term on the LHS), to lowest order it is 
\begin{equation}
\label{eq:botzmann3}
  f_s^1{(\bm{q}})=-{\tau_s(\bm{q})}{{\bm{v}}_s}({\bm{q}}) \cdot \nabla T\frac{{\partial {f_s^0}(\bm{q})}}{{\partial T}} \nonumber,
\end{equation}
or finally
\begin{equation}
\label{eq:botzmann4}
 f_s(\bm{q}) \approx f_s^0(\bm{q})-{\tau_s(\bm{q})}{{\bm{v}}_s}({\bm{q}}) \cdot \nabla T\frac{{\partial {f_s^0}(\bm{q})}}{{\partial T}}.
\end{equation}

Let us now connect the above results (of the stationary case) with the thermal conductivity tensor. The total heat current carried by phonons with single-particle distribution function $f_s{(\bm{q})}$, summing over all different phonon polarizations, is (adjusted to 2D)
\begin{equation}
\label{eq:thermalcurrent}
{{\bm{j}}_Q} = \;\sum\limits_s {\int\limits_{} {\frac{{d^2{\bm{q}}}}{{{{(2\pi )}^2}}}} }    \hbar {\omega _s}({\bm{q}})   {{\bm{v}}_s}({\bm{q}})   {f_s}({\bm{q}}).
\end{equation}
Combining Eqs.\eqref{eq:botzmann4} and \eqref{eq:thermalcurrent} we find (the term containing the equilibrium distribution function of the phonons does not participate in the heat current and is dropped)
\begin{align}
&{\bm{j}}_Q = -\; \sum\limits_s {\int\limits_{} {\frac{{d^2{\bm{q}}}}{{{{(2\pi )}^2}}}} }    \hbar {\omega _s}({\bm{q}})   {{\bm{v}}_s}({\bm{q}})   {\tau _s}({\bm{q}}) \frac{{\partial f_s^0({\bm{q}})}}{{\partial T}}{{\bm{v}}_s}({\bm{q}}) \cdot \nabla T,  
\end{align}
and recalling the definition of the thermal conductivity tensor $\kappa$ via the Fourier law of heat transport (adjusted to a 2D system) which reads
\begin{equation}
{\bm{j}}_Q =- \kappa \nabla T,
\end{equation}
we find for the $\textit{thermal conductivity tensor per unit area}$ the following expression (notice that in order to get the correct units we need to take into account the relaxation time per unit area as defined in Eq.\eqref{eq:relaxtime})
\begin{equation}
\label{eq:thermalconductivity1}
\kappa_{\alpha\beta}= \sum\limits_s {\int\limits_{} {\frac{{d^2{\bm{q}}}}{{{{(2\pi )}^2}}}} }    \hbar {\omega _s}({\bm{q}})   {{v}_s^{\alpha}}({\bm{q}})   v_s^{\beta}({\bm{q}})  {\tau _s}({\bm{q}}) \frac{{\partial f_s^0({\bm{q}})}}{{\partial T}}.
\end{equation}

Before concluding this section, let us mention that all the aforementioned analysis can also be applied to magnons weakly interacting with a phonon bath, as happens in the magnon-dominated transport regime, of course with the appropriate modifications. The more general case in which both types of carriers participate significantly in the total thermal conductivity requires a more sophisticated treatment than the one given here.
\\
\\

\twocolumngrid

\section{\label{sec:appendixc}Technical details for the computation of the line integrals of the various scattering rates}

To calculate the line integrals (reduction comes upon using the property of the Dirac $\delta$ function mentioned in the main text) appearing in various scattering rates (magnonic or phononic), one needs to find the path of integration dictated by the energy conservation constraint. For instance, for the magnon scattering rates of the following general form
\begin{equation}
\left. \frac{1}{\tau_{\lambda}(\bm{k})} \right|_{mp}=\sum_{\lambda'}\int F(\bm{k},\bm{q},T)\delta(\epsilon_{\bm{k\pm q},\lambda'}\pm \hbar\omega_{\bm{q}}\pm \epsilon_{\bm{k},\lambda}) dl(\bm{q}),
\end{equation}
for each given $(k_x,k_y)$ point of interest, one needs to know all $(q_x,q_y)$ points that satisfy the energy conservation constraint $\delta(\epsilon_{\bm{k\pm q},\lambda'}\pm \hbar\omega_{\bm{q}}\pm \epsilon_{\bm{k},\lambda})$ first, and then perform the line-integral over these points numerically.  Due to high non-linearity of the magnon dispersion relations simple analytical expressions are not possible. Thus, the energy constraint was graphically solved like this: For a specific temperature $T$, a grid of $(k_x, k_y)$-points were taken in the vicinity of the various magnon valleys, and for each one of those $\bm{k}$-points a contour plot of the energy constraint was created. From each contour plot, all the $(q_x,q_y)$ points that satisfy the energy constraint for that fixed $(k_x,k_y)$ point were extracted, and were then used to compute the reduced integral. This way, the quantity  $\frac{1}{\tau_{\lambda}(\bm{k},T)}$ for every $(k_x,k_y)$-point was calculated, and further, the whole previous calculation was repeated for each temperature of the chosen temperature window for this study.  In the semiclassical Boltzmann approach to the thermal conductivity, the quantity  $\frac{1}{\tau_{\lambda}(\bm{k},T)}$ enters within a second integral, this time over the magnon momentum space of interest (i.e. over the $\bm{k}$ space), whereby one finally gets the magnon thermal conductivity. A similar procedure is followed for the calculation of the phononic thermal conductivity.

As a last note, to get the constant energy surfaces for a fixed temperature value, for a given wavevector of the one quasiparticle type, the various points of the contour plot of the energy constraint were extracted using the following 'mathematica' command: 
$\texttt{List=Cases[Normal[ContourPlot\_pic],Line[Data\_]} \rightarrow \texttt{Data,5]}$.


%
\end{document}